\documentstyle[11pt,newpasp,psfig,epsf,simplemargins,picinpar,graphicx,eso-pic]{article}

\makeatletter  % Allow the use of @ in command names
\def\fnum@figure{Picture \thefigure.}
\makeatother

\makeatletter
\newcommand\BackgroundPicture[2]{%
  \setlength{\unitlength}{1pt}%
  default \put(0,\strip@pt\paperheight){%
  \parbox[t][\paperheight]{\paperwidth}{%
    \vfill
    \centering\includegraphics[angle=#2]{#1}
    \vfill
}}} %
\makeatother   % Cancel the effect of \makeatletter

\begin{document}
\thispagestyle{empty}
\AddToShipoutPicture*{\BackgroundPicture{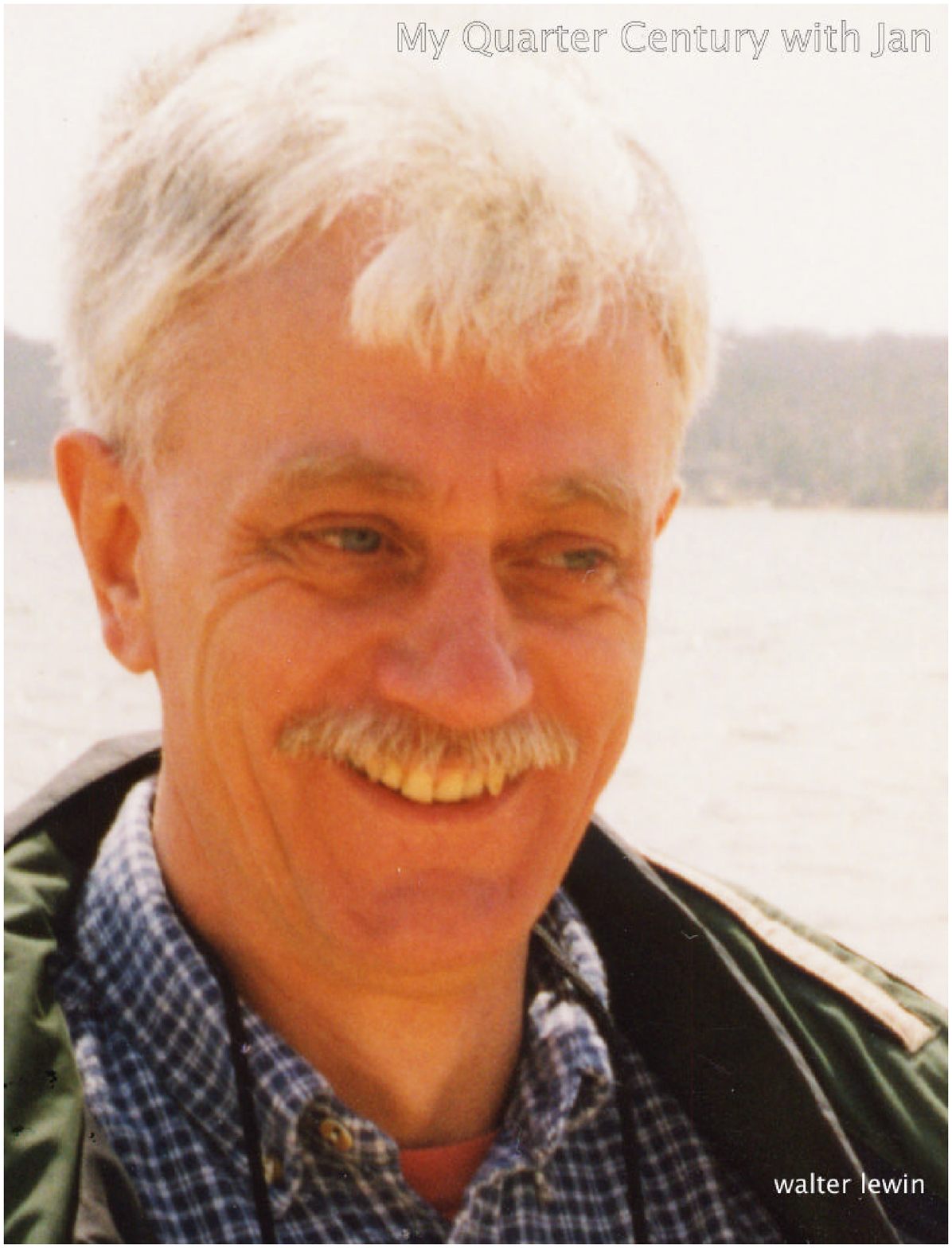}{0}}
\mbox{ }

\newpage

\thispagestyle{empty}

\mbox{ } \vspace{5.7in}

\mbox{ } \hspace{3.2in}{\Large \it For all who loved Jan}

\newpage

\addtocounter{page}{24}
\title{My Quarter Century with Jan}
\author{Walter Lewin}
\affil{Massachusetts Institute of Technology, USA}
\vspace{2cm} 
\noindent A quarter century ago, Jan van Paradijs and I embarked on a scientific
journey, our lives became entwined, and a friendship evolved that
shaped both our lives scientifically and in personal matters.  To tell
our story was very difficult for me. I went through 24 years of notes
in my calendars and re-lived more than 8000 days of my own and Jan's
life, 20 years of slides, 15 years of e-mail. There were times that I
had to stop and leave it alone for a week or more as it became too
emotional. Few people know (perhaps only two) what we meant for each
other. When Jan died, part of me died.

\section{The Early (X-ray Burst) Years}
We first met on April 5, 1976 at ESTEC in Noordwijk, the Netherlands
where I gave a colloquium on X-ray bursts. X-ray bursts were a very
hot topic at the time. This was less than half a year after their
discovery by Josh Grindlay and John Heise with {\it ANS}, and only one
month after the discovery of the Rapid Burster with {\it SAS-3}. The
next year Jan applied for a {\it SAS-3} position at MIT. George Clark
(PI on {\it SAS-3}) asked me to meet with Jan in Amsterdam ``to look
him over.'' After my colloquium at the University of Amsterdam on
February 18, 1977, I had dinner with Jan and some students, among
them Jan's graduate student Ed Zuiderwijk. At that time no burst
source had yet been optically identified, and at dinner I discussed
with them the importance of finding optical identifications. Ed argued
that such identifications would not be of any use.  Jan told me later
that he was very embarrassed about this, and that he was sweating
bricks as he was afraid that this would influence my opinion of him.
He repeatedly kicked Ed under the table in the hope that he would shut
up, but he didn't.

I strongly recommended to George that we invite Jan to join our {\it
SAS-3} group.  Jan shared my contagious enthusiasm, but unlike me, he
had a formal training in astronomy; in that sense he was a {\it real}
astronomer and that was a big plus! All of us at MIT who were doing
research in X-ray astronomy at the time were physicists by
training. The fact that Jan was Dutch may have played a role as well
in my recommendation. I found a copy of George's letter in which he
invited Jan to come to MIT as a visiting scientist; it is dated March
3, 1977.

Jan married Lizzy Soenarjati on April 28, 1977. Lizzy was an
Indonesian woman who worked at the Astronomical Institute in
Amsterdam; they came to the US that September. However, Lizzy soon
lost interest in Jan and she left him within a year. Jan was very down
during this period; he suffered immensely (this was his second
marriage). I spent a lot of time with him in an effort to ease some of
his pains.

Our close scientific collaboration began early in 1978 when X-ray
bursts were still at the focal point of my research. Little did we
know then that we would end up, more than two decades later,
co-authoring 150 publications in refereed journals of which more than
50 related to X-ray bursts.

Following earlier work by Jean Swank and my own group on blackbody
radii of X-ray burst sources, Jan expanded on this in the spring of
1978 by introducing the idea that the peak luminosity in type I X-ray
bursts is a standard candle, and he concluded that the effective
blackbody radius is the same for all burst sources. By assuming that
the standard candle is the Eddington limit of a neutron star, he
derived distances to ten burst sources. This was the first impressive
effort to unify average properties of X-ray burst sources.

In June 1977 Jeff McClintock and co-workers had optically identified
the burst sources 1636--53 and \mbox{1735--44}. That same summer
(1977) I organized a world-wide simultaneous optical/radio/X-ray burst
watch in which {\it SAS-3} observed both sources for several weeks.
Seventeen optical/radio observatories participated, but none detected
a burst from these faint (B$\sim$18) blue stars at the times that {\it
SAS-3} detected X-ray bursts. In the summer of 1978, when Jan was at
MIT, we repeated this, and we succeeded this time. On June 2 Josh
Grindlay and co-workers observed an optical burst from 1735--44
coincident with an X-ray burst that we detected with {\it SAS-3}, and
we made it to the cover of {\it Nature!}

\begin{figure}[!ht]
\plotone{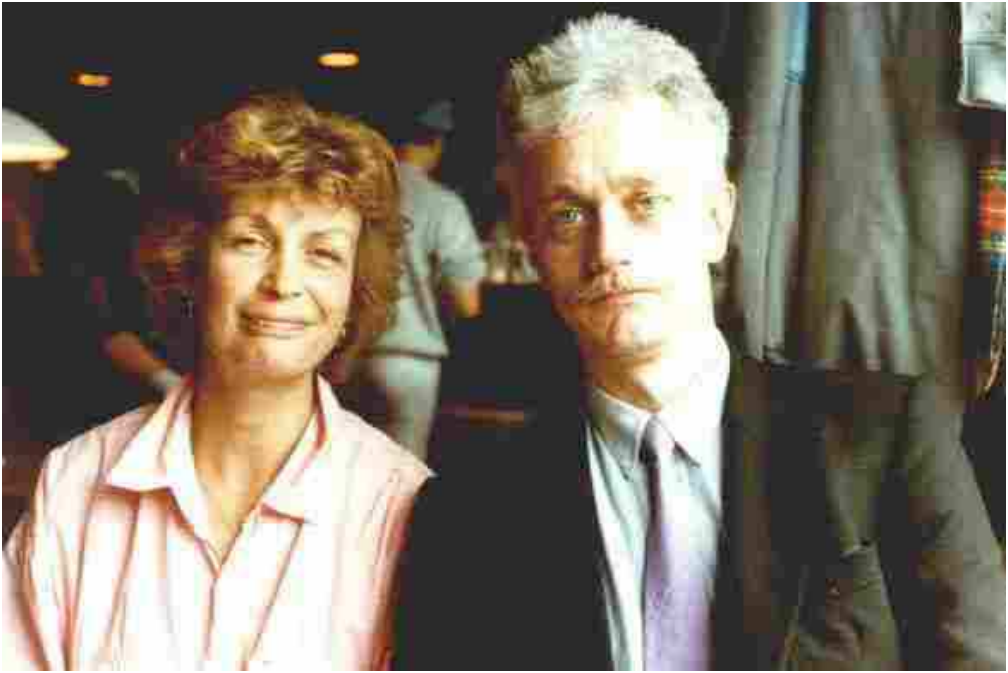}
\caption{Jan and Maria Moesman (1989). This picture was taken by Maria's son
Camiel van Ulden. Courtesy of Maria Moesman.}
\end{figure}
I recall that Jan enthusiastically showed me a letter from Maria
Moesman who had heard about the break up with Lizzy.  Maria was an
employee at the Astronomical Institute in Amsterdam. Her letter was
the catalyst for an on-again off-again relationship that lasted twelve
years. During that time Jan always kept his own apartment in
Amsterdam, first at the Govert Flinckstraat and later (from 1983) at
the Geldersekade. They married on October 26, 1989, bought a house
together on the Twijnderslaan in Haarlem, and divorced two years
later (Picture 1).

During the question and answer period after my talk (on X-ray bursts)
at the Texas symposium on December 19, 1978 in Munich, Boudewijn
Swanenburg pointed out that the gravitational redshift would affect
our estimates of the blackbody radii of burst sources. I had to admit
that we had not taken that into account. Back at MIT I mentioned this
to Jan. He immediately went to work and showed that the X-ray bursts,
at least in principle, contain key information on the mass and radius
of the neutron stars. He published his findings in 1979. In the years
that followed, we developed several methods and did our utmost to
derive ``reliable'' values for mass and radius from burst
spectra. Unfortunately we never achieved an accuracy that made the
results very meaningful. The systematic errors were always the killer.

\begin{figure}[!t]
\plotone{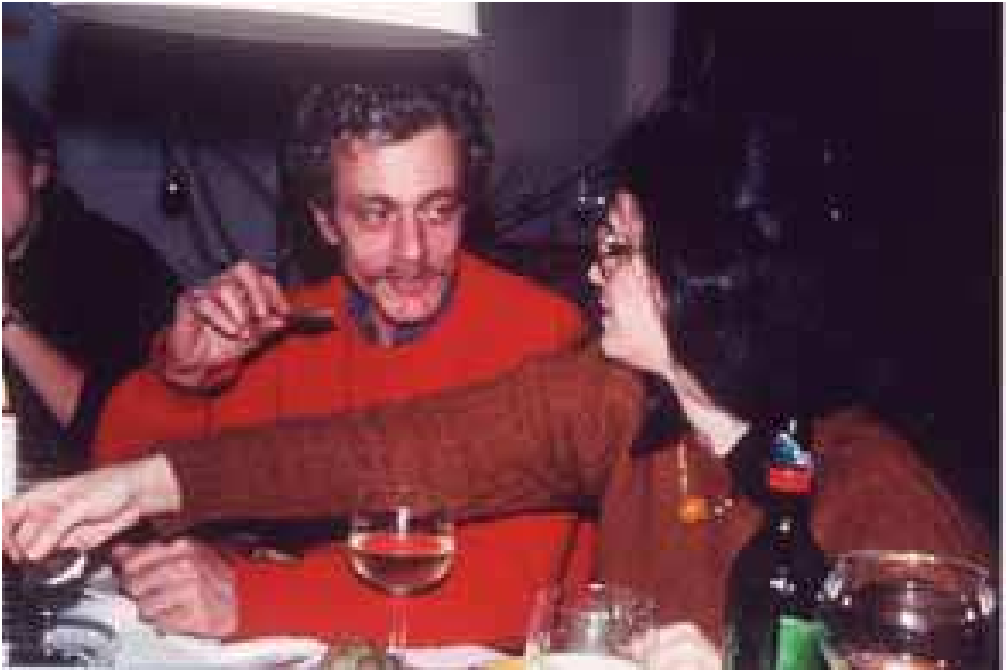}
\caption{Jan was a frequent dinner guest at my apartment in
Cambridge (10 Center Street).}
\end{figure}

During his two years at MIT Jan was a frequent guest at my
apartment in Cambridge (Picture 2) where I lived with Ewa Basinska who
became my second wife in January 1981.  In 1979 he returned to the
University of Amsterdam.  On November 15 of that year, Jan and Maria
attended the Piet Mondriaan Lecture {\it "Wetenschap dat is geen
Kunst"} which I gave in the Sonesta Koepel in Amsterdam.

Together with Holger Pedersen, Jan and I observed for a few weeks in
the summers of 1980, 1981 and 1982 at the European Southern
Observatory (ESO) in Chile. We spent all our observing time on
\mbox{1636--53}.  Simultaneously, the Japanese X-ray Observatory {\it
Hakucho} (which means Swan) was recording the X-ray bursts. It was
really Holger who did most of the work, and who deserves most of the
credit. We detected dozens of optical bursts, many of which were also
seen with {\it Hakucho}. Our observations for the first time covered a
wide range of optical burst fluences and peak fluxes. We also
discovered the orbital period of 1636--53 in the optical data.  Those
long nights at the telescope were very special. To pass the time, the
three of us would frequently sing a song together after a 2-minute
integration (independently we monitored the data continuously with a
time resolution of 20 msec). It was a song that we had made up
ourselves; it had two versions depending on whether we measured an
increase
\begin{figure}[!ht]
\plotone{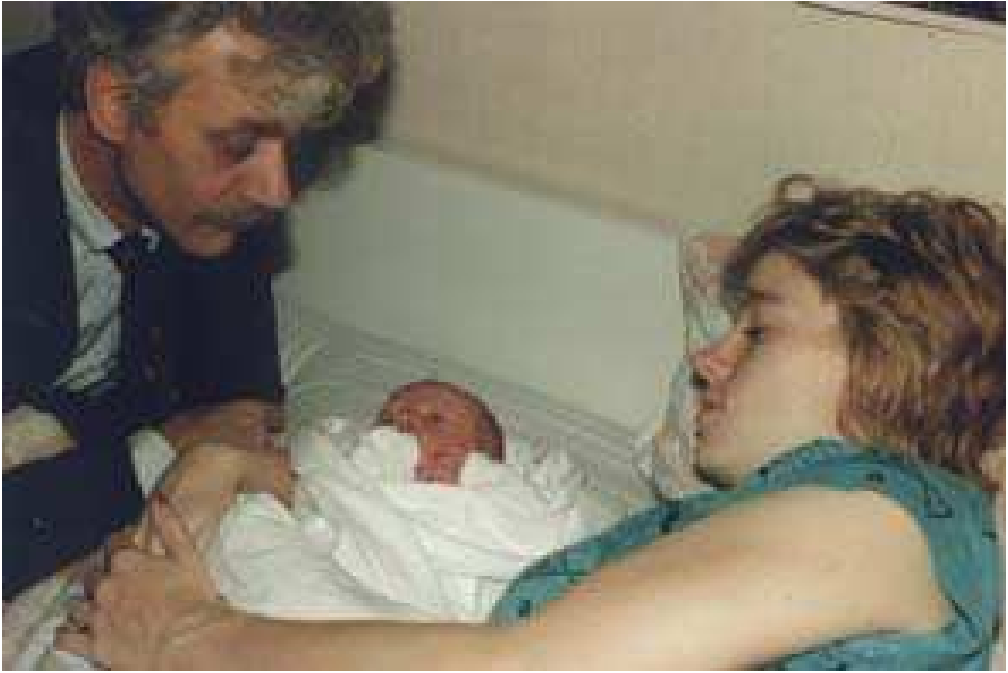}
\caption{Jan with his daughter Deirdre shortly after the
birth of his first grandchild Quinten (Sept.~1989). Picture courtesy
of Maria Moesman.}
\end{figure}
or decrease in the accumulated counts.  If anyone had walked in on us
(s)he would have thought that we were lunatics (perhaps we were). Jan
and I later in life often sang that song together. Those were such
happy times for all three of us. I am sure that Holger too will never
forget them.  Writing about it now, two decades later, is very
difficult; it hurts.
 
In July 1983, Joachim Tr\"{u}mper, Jan and I shared an apartment
during a two-week meeting on ``High Energy Transients in
Astrophysics'' in Santa Cruz. There was a nearby nude beach where we
often went; I could even show some pictures!

\begin{figure}[!t]
\plotone{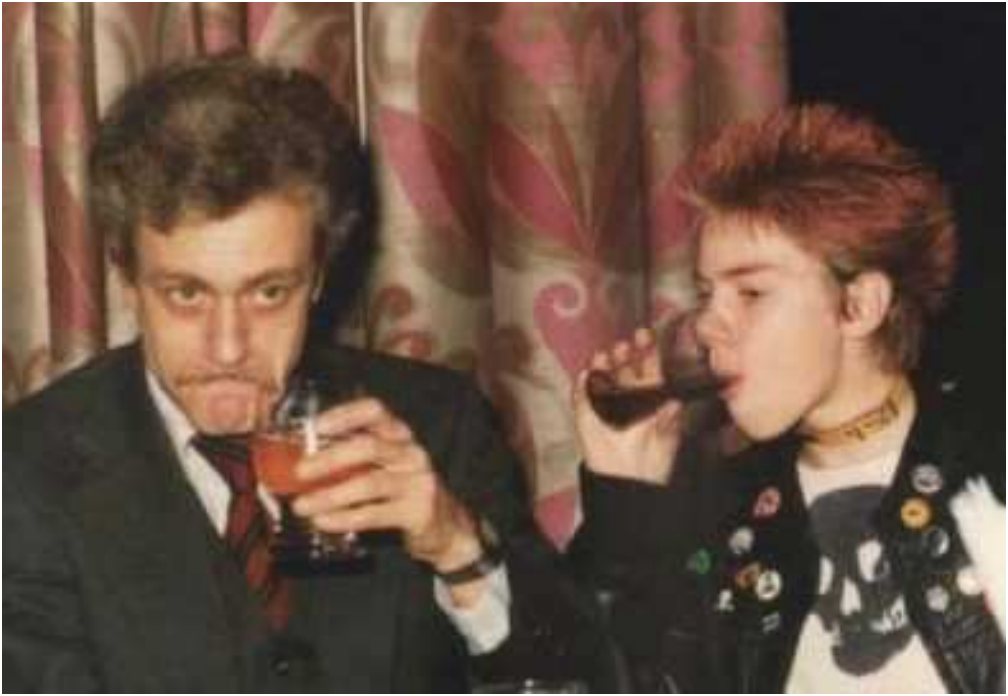}
\caption{Jan and his son Oskar at the 35th anniversary of Jan's
parents (1981). Picture courtesy of Maria Moesman.}
\end{figure}

In 1983 Jan was awarded the Schmeitz Prize for Astronomy of the
``Nederlandse Astronomen Club'' (awarded every three years for the
best astronomical work done by a Dutch astronomer younger than 40).  I
attended the ceremony in the Netherlands on January 4, 1984, and that
is where I met for the first time Jan's daughter, Deirdre, from his
first marriage with Kitty (whom I never met). Over the years I have
met Jan's son, Oskar, countless times, but I do not remember whether
he was present then (see Pictures 3 and 4). On this very happy and
special occasion for all of us, Jan's mother, who was in the hospital,
gave Jan roses which he always kept (see Picture 24).  She died two
months later on February 26.

\section{The EXOSAT (QPO) Era} 
During a brainstorm in the fall of 1983, when Jan, Joachim Tr\"{u}mper
and I talked about all kinds of exciting science that we might want to
do with \mbox{{\it EXOSAT}}, I suggested to observe several of the
brightest low--mass--X-ray binaries (LMXBs) in an attempt to discover
their millisecond spin periods. In 1982 Ali Alpar and co--workers had
persuasively suggested that LMXBs are the precursors of millisecond
radio pulsars; during the mass transfer phase the neutron stars spin
up. \mbox{Co-I} Michiel van der Klis was present in Darmstadt when our
\mbox{GX 5--1} observations were made on September 18, 1984. Michiel,
together with Fred Jansen, skillfully did the data analysis. No spin
periods were found, but they discovered the intensity dependent
quasi--periodic--oscillations (QPO) in the range 20--40 Hz. This
discovery triggered a wealth of activity (both in theory and in
observations) in which Michiel evolved as the world's leader in X-ray
timing which he still is to date. For this, he received the Rossi
Prize in 1987.

\begin{figure}[!b]
\plotone{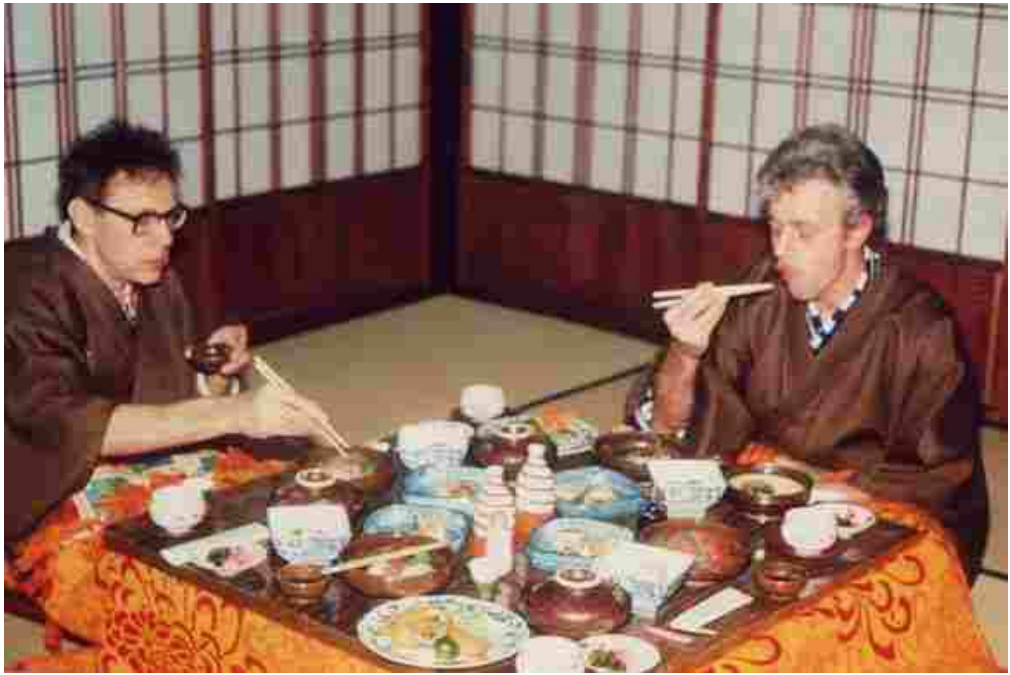}
\caption{Dinner in Kyoto with Jan (1985).}
\end{figure}
I spent all of calendar year 1985 (as an Alexander von Humboldt
awardee) at the Max-Planck Institut f\"{u}r Extraterrestrische Physik
(MPE) in Garching and Jan often visited me there for several weeks at
a time.  The year started out with an X-ray astronomy meeting in
January in Tokyo of which Yasuo Tanaka and I were the
organizers. After the meeting Jan, Ewa, Joachim Tr\"{u}mper and I
stayed in Japan for an extra week of vacation. We were treated
wonderfully by our hosts Sigenori Miyamoto in Osaka (Picture 9) and
Hideyo Kunieda in Nagoya.  We also visited Nara and Kyoto where we
attended a 6--hr long Noh performance. This was our first exposure to
Noh theater; Jan and I both loved it. It was one of my best vacations
ever! I show here five pictures (5--9) of Jan during that wonderful trip.

\begin{figure}[!t]
\plotfiddle{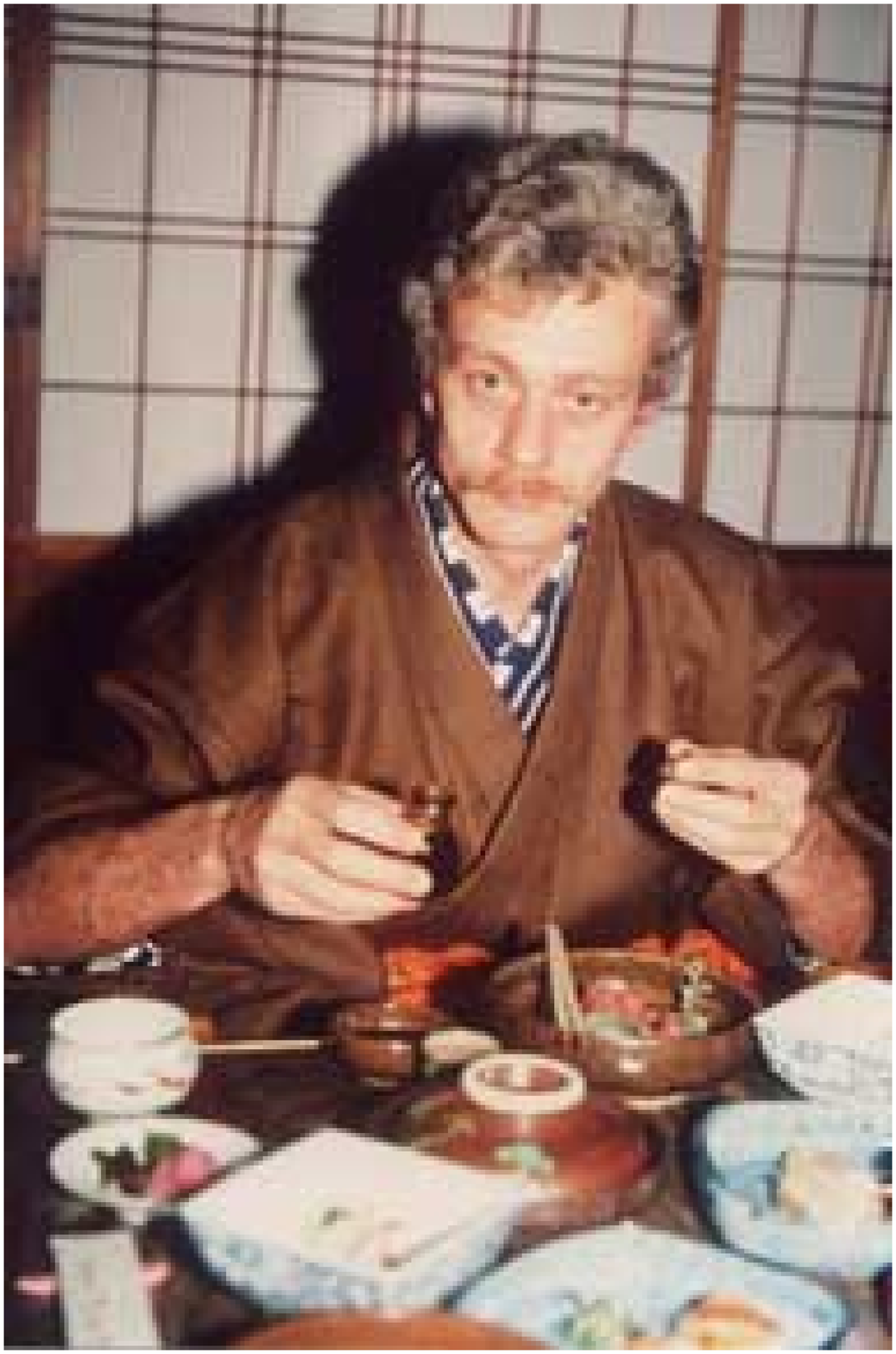}{4.9in}{0}{45}{45}{-125}{-10}
\caption{Jan in Kyoto.}
\end{figure}

There was an X-ray meeting in Bamberg, Germany April 17--19, 1985.
Michiel van der Klis was scheduled to present our GX 5--1 QPO data on
the 18th; this would have been the highlight of the meeting. Jan and I
were awaiting Michiel's arrival, but he did not show up. We made
several calls but no one in Amsterdam knew what had happened. It
turned out only later that Michiel's flight from Schiphol had been
canceled. We decided to go ahead anyhow. Jan gave the talk without any
authentic plots. We made some rough sketches on a blackboard, the way
we remembered things; we both were very familiar with the results.  On
March 8, I had given a colloquium on this at MPE, but I had left all my
plots in Garching.

Michiel (at ESTEC in Noordwijk), Jan and I (at MPE in Garching), wrote
the paper on the intensity-dependent QPO in GX 5--1.  It was very
intense work, with many fax exchanges between MPE and ESTEC. When we
were finished, Jan and I were so proud and happy that we asked Ewa to
take our picture
\begin{figure}[!b]
\plotone{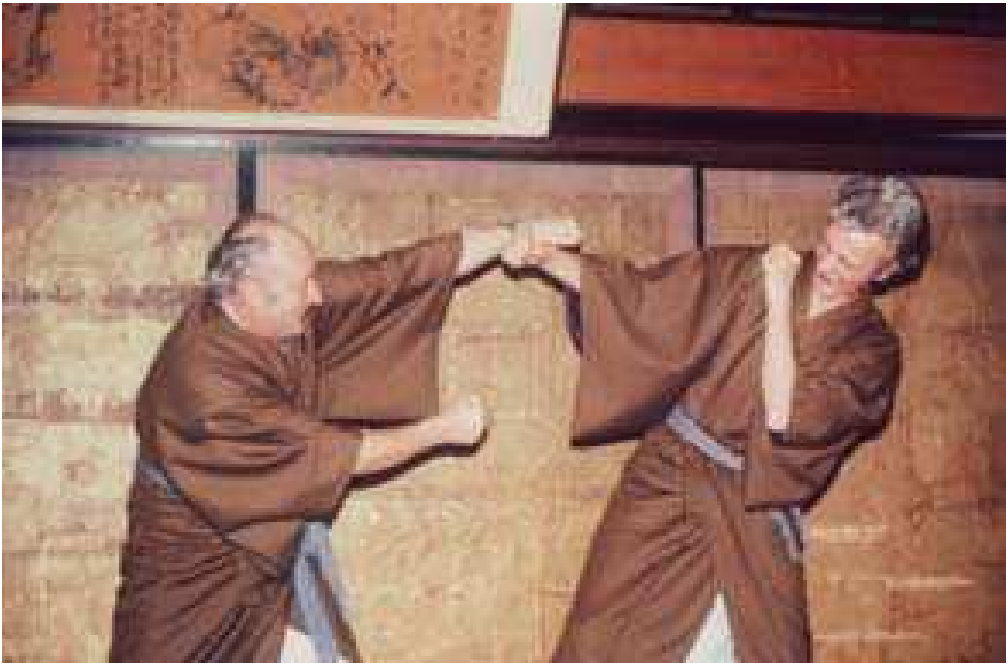}
\caption{A relaxing boxing match with Joachim in Kyoto
before dinner.}
\end{figure}
\begin{figure}[!t]
\plotone{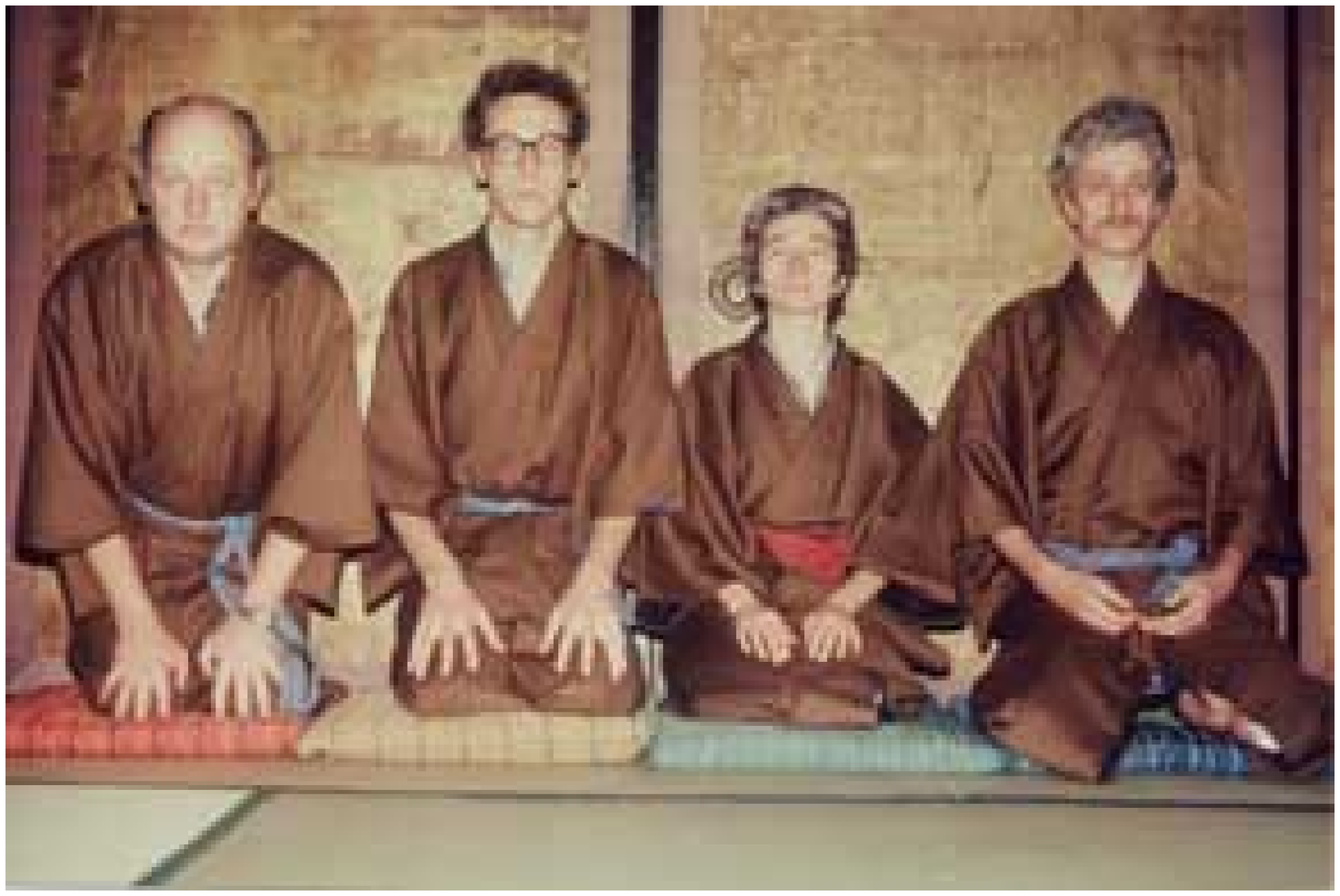}
\caption{Meditation before dinner in Kyoto. From left to
right Joachim, Walter, Ewa and Jan.}
\end{figure}
in our apartment in Garching while we were holding the manuscript in
our hands (Picture 10).  Michiel had to be in London on May 10 to
present the new results to the Royal Astronomical Society, and he
hand-delivered our manuscript to {\it Nature}. On May 10 Jan and I
visited the Lenbach Haus in Munich which holds a fabulous Kandinsky
collection. The next day, we visited the ``Russenhaus'' in Murnau
(Southern Bavaria) which is where Kandinsky and Gabriele M\"{u}nter
lived from 1909--1914.

\begin{figure}[!b]
\plotone{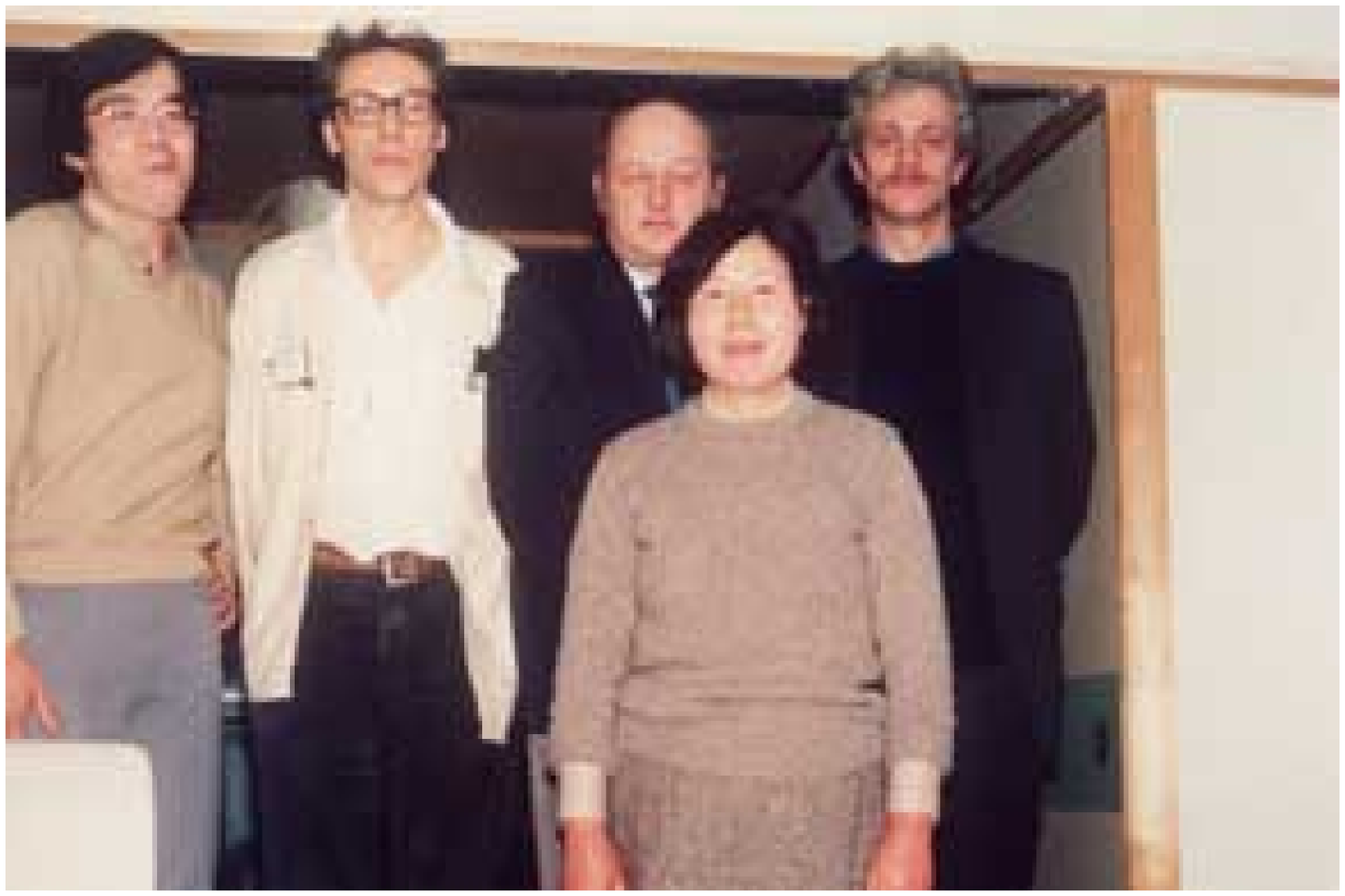}
\caption{A visit to the home of Sigenori Miyamoto in Osaka
(1985). From left to right: Hiroshi Tsunemi, Walter, Joachim (tired?),
Mrs.\ Miyamoto, Jan.  We were standing in the doorway that connects
two rooms.  Notice that Hiroshi, Jan and I were a bit too tall!}
\end{figure}

Jan spent again a few weeks in Garching in August, October and
November (1985). The August visit was the most eventful. We went to
the operation center (ESOC) in Darmstadt as we often did when one of
our approved {\it EXOSAT} observations was made. During that August
visit, one of our targets 1728--34 (also called the slow burster) was
going to be observed.  1728--34 is only 0.5 degree from the Rapid
Burster (my pet source). With 1728 smack in the middle of the field of
view, the Rapid Burster (RB) is also in the field of view, and Jan and
I were hoping that the RB would be active as we were much more
interested in observing the RB in outburst than in 1728. Before the
1728 observations started, we talked to Nick White (who was in charge)
and asked his approval to move to the RB in case it was active. Nick
agreed and went home.  Luigi Stella was the duty scientist. The RB
turned out to be active, and Jan and I were delighted! However, we
were not aware of a rule that TOO's (targets of opportunity) that
would be accidentally discovered were the property of the duty
scientists at ESOC. Luigi called Nick who came back to the operation
center, and Jan and I were told that our 1728 observations were going
to be postponed to make room for the RB observations which would be
the property of the ESOC duty scientists.  Jan and I were very upset,
as it was our understanding from talking to Nick before he left, that
Jan and I would be allowed to observe the RB.  I do not remember all the
details. Neither does Nick or Luigi who both have kindly helped me to
reconstruct what happened, and who did what, when and where.

Nick had to go to ESTEC (Noordwijk) the next day, and discussed the
situation with Tony Peacock.  In Darmstadt, Jan and I received a telex
from Tony: ``The Rapid Burster is not 1728--34.'' The message was loud
and clear; we would not have any RB data rights. I wrote in my
calendar on August 28: {\it call from Tony Peacock \ldots poor Jan!}
Tony called us in Darmstadt (while Nick was with Tony in his office),
and he lashed out at Jan with a barrage of never-ending
accusations. Jan was dumbfounded and very upset.  However, in a way
this cleared the air. When a few days later the storm had settled a
bit, we were offered co-ownership of the RB data with the
understanding that Luigi would do the analysis and be lead author on
the publications.  The scientific results that came out of those RB
observations were fabulous (intriguing QPO behavior, still not
explained as of this date), and Jan and I were content with this
compromise.

{\it EXOSAT}, and that certainly includes Tony, Nick and Luigi,
treated us very well. Jan and I co-authored more than 60 refereed {\it
EXOSAT}-related papers; we (with Michiel van der Klis) were granted
hundreds of hours of observing time to study burst sources and QPO in
low--mass X-ray binaries; Tony even granted us discretionary time to
observe the illusive {\it Perseus Flasher}; it turned out later that
the light flashes, observed by amateur astronomers, were due to
reflections from satellites. The {\it EXOSAT} era was unquestionably a
highlight in our lives.

In 1987 Jan surprised me on my birthday with a very large photograph
(poster size) of the time-evolved QPO from a portion of the Rapid
Burster data. I framed it; it is very abstract, and it has merged
seamlessly with my contemporary art collection.

\begin{figure}[!t]
\plotfiddle{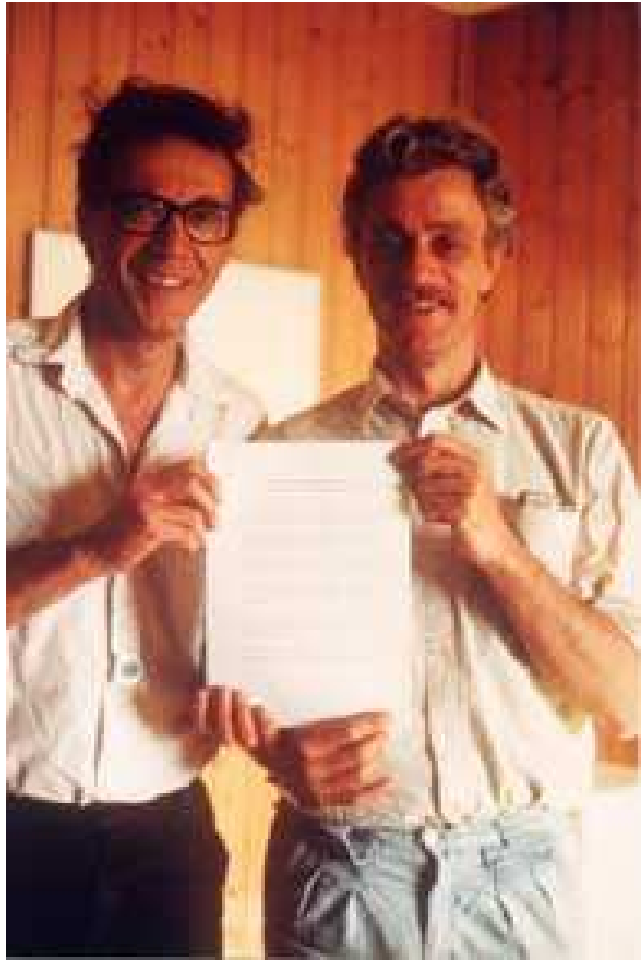}{4.9in}{0}{45}{45}{-125}{-10}
\caption{When our GX 5--1 QPO manuscript for {\it Nature} 
was finished (see
text), Jan and I were very happy, and we proudly displayed it
in my apartment in Garching where I lived all of
calendar year 1985.}
\end{figure}

In October (1985) Jan joined Ewa and me for a few days in our favorite
Alpen Gasthoff Bergheim in Fotschertal near Sellrain (Tirol, Austria)
where we hiked (Picture 11).  Ewa and I were fanatic mushroom
hunters. I kept a diary in Bergheim where we were frequent guests. In
that diary I wrote down many details about our visits, including the
locations where we found our favorite mushrooms (chanterelles and
boletuses). However, I did not want any of the other guests (or the
owners of Bergheim, Sissy and Peter for that matter) to know where we
found them as they too were devoted mushroom hunters who also kept the
locations of their finds very secret. I therefore introduced code names
for the locations of our finds.  I named one of the logging roads the
{\it Jan van Paradijs Road} (Jan loved it when I told him).

Jan and I attended the April 1986 meeting in Tenerife, and in August
we worked together for a week in Garching. Jan made several trips to
MIT that year and in 1987. I spent ten days in Amsterdam in June of
1988 where we, with Maria and Oskar, celebrated Jan's 42nd birthday
(June 9) at his apartment on the Geldersekade.  Jan became professor
in Amsterdam in January 1988. In September that year, we both gave
several lectures at the summer school in Erice, Sicily on ``Neutron
Stars and their Birth Events'' which was
\begin{figure}[!t]
\plotfiddle{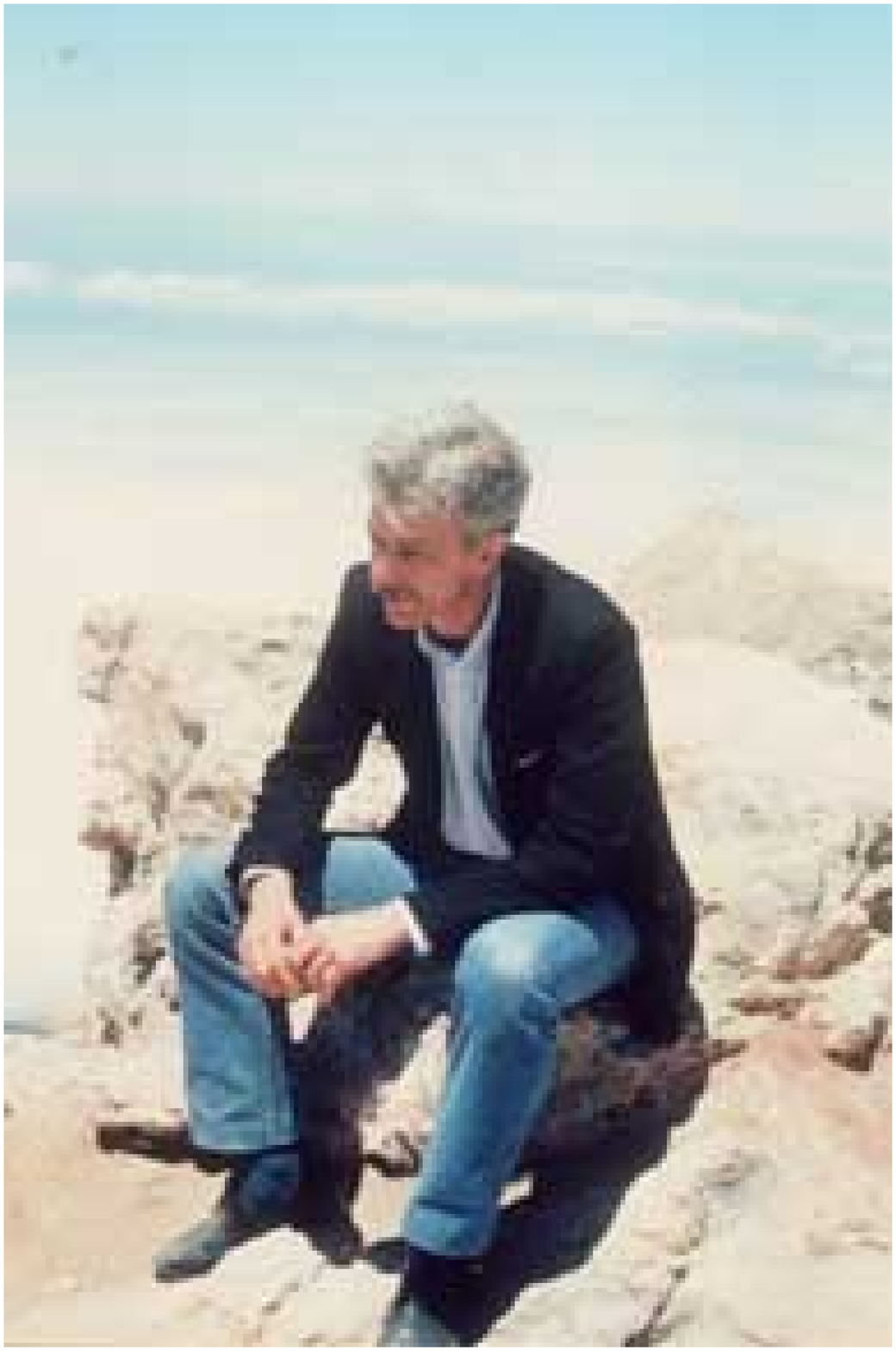}{4.9in}{0}{45}{45}{-125}{-10}
\caption{Jan resting during one of our hikes in Fotschertal, 
Austria (1985).}
\end{figure}
organized by Wolfgang Kundt. Afterwards I went back with Jan to
Amsterdam where I attended one of his classes (he taught a course on
stellar atmospheres) at the Astronomical Institute which was then
located at Roeterstraat 15 (the Institute moved to the Kruislaan in
January 1991).

In 1989 when Jan and I discussed the kind of things we wanted to do
with {\it ROSAT}, Jan suggested that we make optical observations of
the Andromeda Nebula (M31) using the Michigan-Dartmouth-MIT
observatory at Kitt Peak, and cover the same regions (about one square
degree) that Joachim Tr\"{u}mper and co-workers were planning to
observe with {\it ROSAT}.  We approached Joachim who liked the idea
very much. My graduate student Eugene Magnier made the observations in
the fall of 1990 and 1991.  This enormous undertaking resulted in an
optical catalogue of 361,281 stars in {\it B,V,R,I} to typical
completion limits of about 22-21 magnitude, and it led to the
discovery of many new OB associations.

I stayed with Jan and Maria in Haarlem in June of 1990, and again in
December. Though we were not religious, on X-mas eve we attended the
mass in the St.\ Bavo Basiliek in Haarlem. We had a splendid party on
the 25th (there were at least 15 people). Maria had made my favorite
course: {\it huzarensla} (this cannot be translated). It was in this
year that I noticed a very interesting work of art in Jan's living
room at the Geldersekade (see Picture 25). It was made by his brother
Ren\'{e}. I was very impressed and I bought two works of art from
Ren\'{e} (a few years later I would buy three more). By this time, Jan
had met Chryssa Kouveliotou (a Greek astrophysicist).

In 1991 we attended a workshop in Santa Barbara at the Institute of
\mbox{Theoretical} Physics (ITP). We went twice for 2 weeks (in Febr/March)
and in May (prior to the AAS meeting in Seattle). Our February/March
visit was almost entirely devoted to writing proposals.  This was not
quite the purpose of that workshop, but it so happened that an
important proposal deadline was coming up.  We needed access to a fax
machine during evenings when the administration was closed and when
the fax machine was behind locked doors. Jan and I both had 24 hour
access at our home institutions to fax machines, and we were very
unhappy that we were not allowed to use the ITP fax machine after
hours. Ed van den Heuvel (one of the organizers of the workshop) came
up with a brilliant solution: we rented a fax machine which was
installed in our office at ITP, and we met our deadlines.  Not a bad
solution for a theoretician!

I recall that we made several naked-eye observations of sun spots in
Santa Barbara.  We would wait till the sun had almost set when it had
dimmed and turned beautifully dark red. Jan had somehow injured his
knee just prior to coming to Santa Barbara, and he was on crutches
part of the time.  That same year (1991), we attended the X-ray
meeting in Garching (October 7--9) which Joachim Tr\"{u}mper, Wolfgang
Brinkmann and I had organized.

\begin{figure}[!b]
\plotone{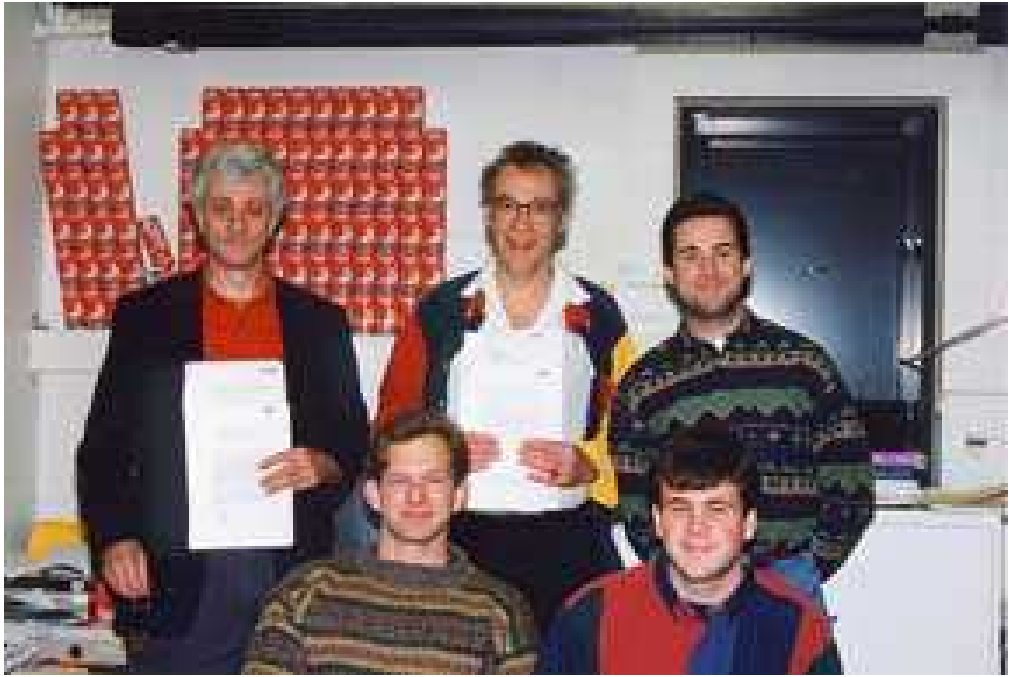}
\caption{At MIT (January 1995). Jan and I are proudly
holding up the final galley proofs of our chapters in {\it ``X-ray
Binaries.''}  We are accompanied by my graduate students Bob Rutledge
(back row) who composed the Coca Cola sculpture, Derek Fox (left front
row), and Jefferson Kommers.}
\end{figure}

During the summer of 1992, I spent three months at the Astronomical
Institute in Amsterdam (sabbatical) and lived in Jan's apartment at
the Geldersekade while Jan was in Huntsville, AL, visiting Chryssa who
worked at the Marshall Space Flight Center. They married in the
Netherlands on October 27 (this was Jan's fourth marriage).

A highlight in 1993 was a visit to NASA's swimming pool in Huntsville.
I visited Jan and gave a colloquium at Marshall Space Flight Center on
multi-wavelength observations of the so called Z-sources. My friend
and ex-collaborator Jeff Hoffman was also in town. He and several
other astronauts were in training for the {\it Hubble} repair
mission. The huge swimming pool was extensively used to simulate
weightlessness. On July 6, Jeff gave us a delightful tour of the
\mbox{facilities} while the training was going on.  Later that year in
September, Jan and I attended a meeting in Turkey which was organized
by Ali Alpar. It was a great meeting. I recall an excursion to the
town of Myra where we sang Dutch Sinterklaas songs (Jan was the
conductor) in the ruins of the church where St.\ Nicholas (the bishop
of Myra) frequently said mass, and where he was buried in the 4th
century AD. Henk Spruit video-taped this hilarious event.

\begin{figure}[!t]
\plotone{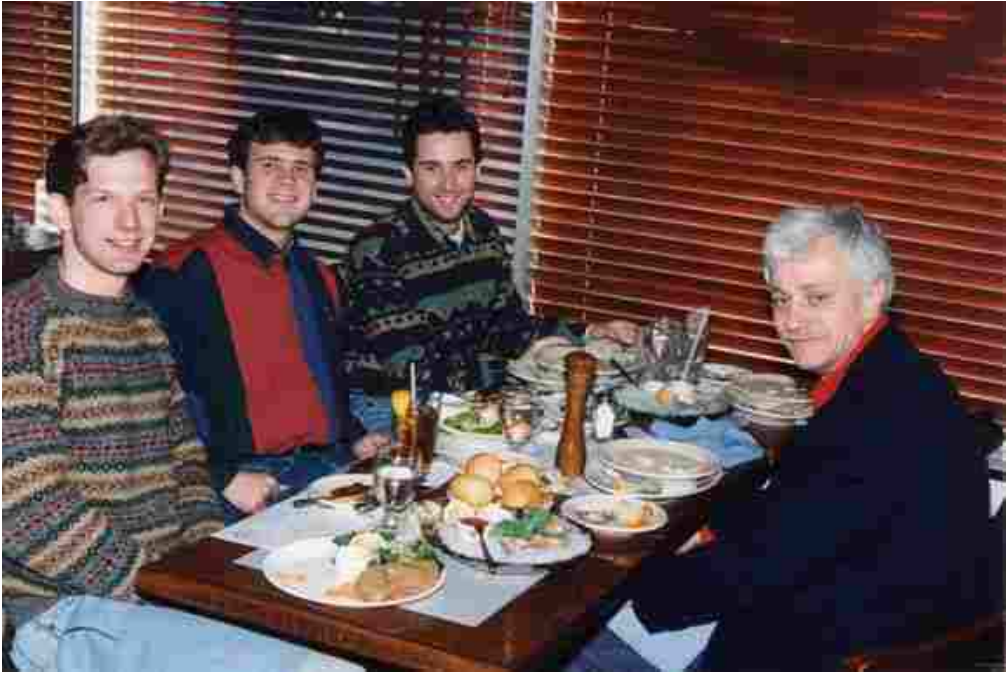}
\caption{Lunch at Legal Seafood near MIT (January
1995): smoked salmon, raw cherry stones, and Legal
Seafood's world famous clam chowder! From left to right: Derek Fox, 
Jefferson Kommers, Bob Rutledge, Jan.}
\end{figure}

In 1994, Jan became the Pei-Ling Chan Professor at the University of
Alabama, Huntsville (UAH). He divided his time between Amsterdam
(June-December) and Huntsville (January-May).  While in Huntsville,
Jan worked mostly at Marshall Space Flight Center.  The two
\mbox{part-time} positions had pros and cons. There were many
plusses. He could spend half the year with Chryssa and he could also
have time to himself while in Amsterdam. He very much cherished that
as he needed a lot of time alone.  Without it, to use his own words,
{\it zou ik knettergek worden} (I'd go nuts).  In addition, his
contact with the group at Marshall spurred his interest in gamma-ray
bursts, and this led in 1997 to a great triumph for Jan (see below)!
\mbox{\hspace{.2in}}

\begin{figure}[!b]
\plotone{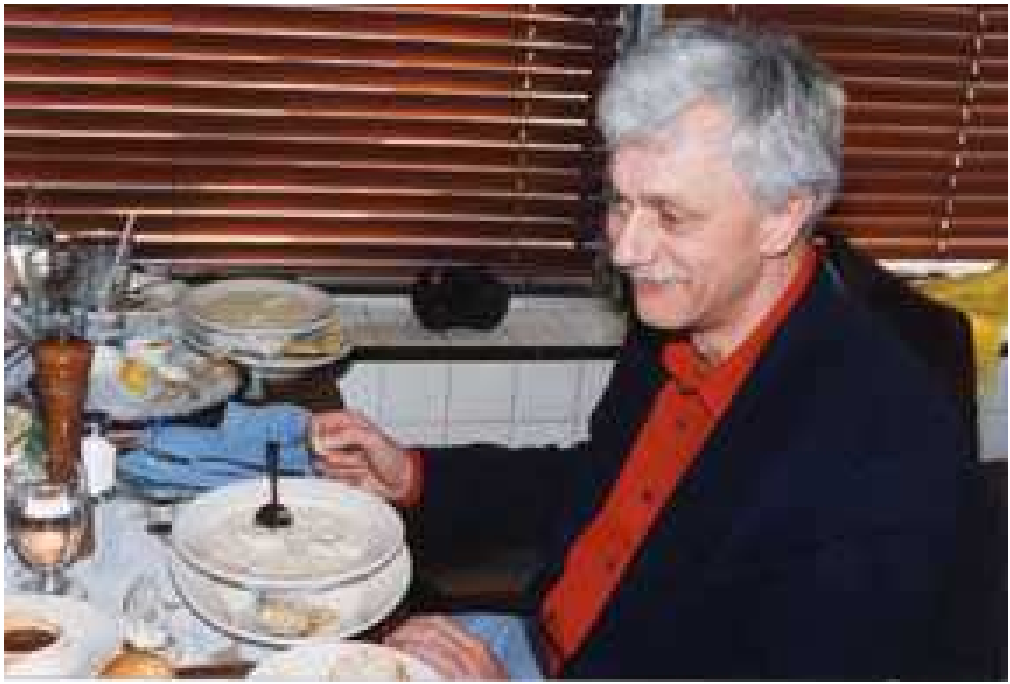}
\caption{Jan loved Legal Seafood's famous clam chowder, and 
who doesn't?}
\end{figure}

There were negatives as well.  Jan did not feel well integrated in the
group at Marshall. He was very good and fast in interpreting results
(he loved that), but he felt that his eagerness to contribute was not
always appreciated.  He also felt that his half-time position in
Amsterdam barred him from certain prestigious positions in Holland
that he would have liked and that, he believed, probably would have
come his way had he continued to spend full time there.

In January 1995 Jan and I celebrated my 59th birthday in Cambridge. We
finished the galley proofs of our chapters in {\it X-ray Binaries},
the book that we had worked on for more than two years (Pictures
12-15). On March 28 we were both members of Erik Kuulkers'
\begin{figure}[!t]
\plotfiddle{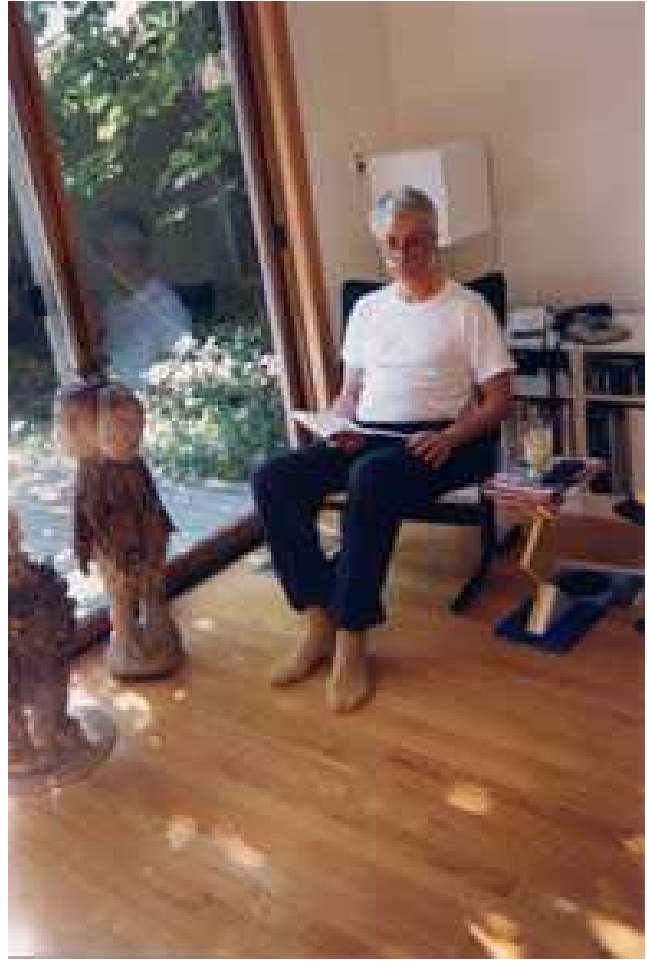}{4.9in}{0}{45}{45}{-125}{-10}
\caption{Jan, still in pyjamas, early morning at my home
(139 Cushing Street) in Cambridge, MA (1995).  This was a very typical
pose for Jan; nothing was more relaxing for him than reading.}
\end{figure}
PhD committee. In Holland, the PhD defense is a wonderful
ceremony. Professors dress up in monkey suits (Picture 16), and the
candidate has to answer questions (for 45 minutes) from the committee
members in the presence of the candidate's parents, friends and loved
ones. In November that year, I spent ten days with Jan in Amsterdam
where I gave a talk at the symposium in honor of Ed van den Heuvel's
very prestigious Spinoza Prize.

Jan got very ill with pneumonia in April 1996. The doctors in
Huntsville were treating it unsuccessfully, and he almost died.  Then
they did tests and decided it was {\it Klebsiella pneumonia}.  They
were able to treat him correctly in the nick of time.  However, the
tests revealed polyps in his colon which, when removed, proved to be
malignant.  The irony was that the unusual pneumonia that almost took
his life ended up saving his life.  There is a good chance that if the
doctors hadn't flubbed the diagnosis to begin with, the polyps would
have gone undetected and his colon cancer might have been caught at a
much later, fatal stage.  All this was extremely worrisome as cancer
runs in Jan's family. Shortly after his recovery, Jan and Chryssa
visited me in May (Picture 17). We walked on the beaches of Plum
Island (where Jan got many nasty insect bites), and we visited the
Fruitlands Museum in Harvard, MA, an hour's drive from Cambridge.

\begin{figure}[!b]
\plotone{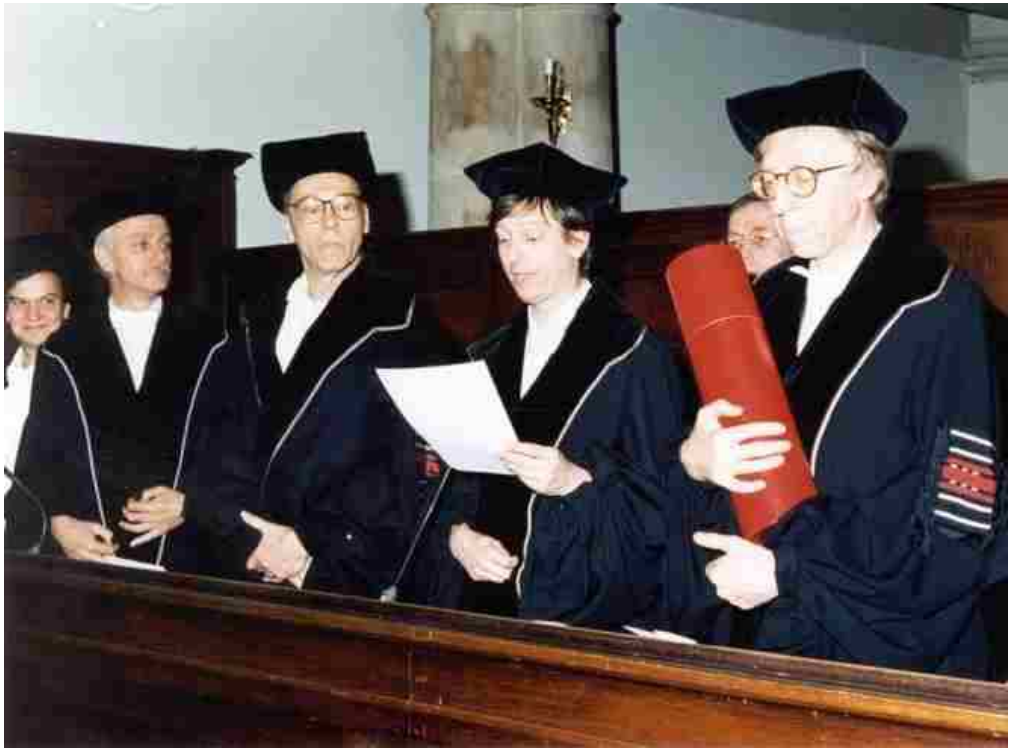}
\caption{During the PhD defense ceremony of Erik Kuulkers in Amsterdam
(March 28, 1995).  From Left to right: Frank Verbunt, Jan, Walter,
Michiel van der Klis, Wim Hermsen (in the back), Ed van den
Heuvel. Photo courtesy Erik Kuulkers.}
\end{figure}

During the summer of 1996, prior to the COSPAR meeting in Birmingham,
I was Jan's guest in Amsterdam. I recall a visit to the Stedelijk
Museum and a very nice dinner with Oskar.  Jan knew little about
contemporary art. I would therefore not miss an opportunity to take
him to art museums and expose him to it. Jan was always willing to
explore something new, and even though contemporary art did not have
his immediate interest (he had so many other interests), he would
never ridicule it (as so many other ignorant people do).  Jan and I
attended the COSPAR meeting, and we met Chryssa in Birmingham. After
the meeting the three of us toured through Wales.  In November that
year, we organized a two-day meeting at MIT on GRO~J1744--28 (the
famous bursting pulsar).

\begin{figure}[!t]
\plotone{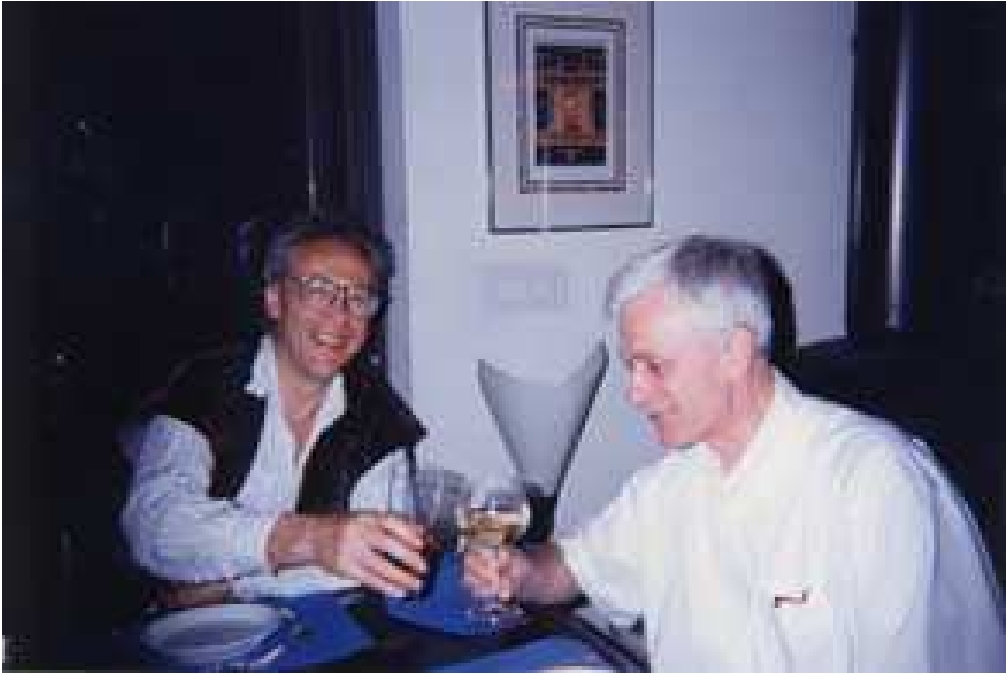}
\caption{Dinner with Jan in Cambridge (May 1996).}
\end{figure}

\section{Jan's Many Interests} 
Jan had many interests outside science.  Collecting books was his
major hobby. There was almost no room in his apartment at the
Geldersekade to hang any pictures on the walls as they were almost
completely covered with book shelves (Pictures 21, 24, 25). One room
was virtually filled with astronomical journals.  Jan loved to browse
through antiquarian book stores and markets. In the evening after work
he would often relax by drinking a beer, reading in one of his recent
acquisitions. There were always piles of books on the floor and on his
desk.

In 1989 Jan discovered the Times Literary Supplement at the house of
an architectural historian in Amsterdam, and became an avid subscriber
and reader of that journal.  Jan would collect the 52 issues each year
and carry them to Athens where a bookbinder bound them to his
specifications.  The journals traveled from their point of origin in
England to Huntsville to Athens to Amsterdam, arriving at their final
destination in blue covers with the year stamped in gold.  He also
collected Prisma Books which covered a wide range of subjects: novels
{\it (Het geheime wapen van tante Searwood)}, stories, poems, art,
dictionaries, biographies {\it (Lodewijk XV; Hendrik VIII)}, biology
{\it (Het leven der mieren)}, science {\it (Van atoom tot heelal)},
health {\it (Word slank en blijf gezond)}, history, camping, crafts
{\it (Prisma-breiboek)}.  He did not read them, he simply wanted to
achieve the goal of getting the first 500 books (Picture 24). He came
very close; I believe he only missed about a dozen. He always carried
with him a list of the missing numbers, and whenever we passed by book
markets (e.g.\ Oudemanhuispoort in Amsterdam) he would look for those
that were missing. The price for one such book was about one Dutch
guilder (4 dimes).

Jan also collected the daily comic about Tom Poes and Heer Bommel by
Marten Toonder. Every Dutchman knows about Tom Poes and Heer Bommel
whose satirical social commentary and word play are closest to Walt
Kelly's Pogo in the US. I had strict instructions: when staying at his
place in his absence, I was allowed to throw out his newspapers (NRC)
but I always had to cut out the above comics and save them for him. He
also had a much admired collection of books featuring the adventures
of Erik de Noorman, a comic strip he had read in the newspapers as a
child in Haarlem. These old books have become collector's items.

Jan started to collect labels of beer bottles in the nineties (Picture
19).  Removing the label from a bottle gave him great pleasure, and he
had developed special techniques (which he once showed me in detail)
on how to remove them without damage.  He explained to me that there
are various ways that the labels can be glued on the bottle, and the
``right'' removal technique had to be chosen accordingly. Processed
labels were flattened in a huge dictionary before he transferred them
to the pages of his album.

Jan's intellectual interests found an outlet in a number of
activities.  He was invited to join the Studio Gaudioque (SGQ), a club
of professors from the University of Amsterdam who met monthly in
their homes to discuss their own research in a variety of fields.  He
was on the editorial board of the University of Amsterdam book review
periodical De Amsterdamse Boekengids.  When he was invited to join the
venerable Hollandsche Maatschappij der Wetenschappen he added a series
of their publications to his library.  He took a special pleasure in
his membership because the society is located in his home town of
Haarlem.  At the dinner parties he gave (Jan could cook rabbit very
well!), conversations ranged over many topics: politics, history,
epistemology. He found common ground with historians because, as he
pointed out, astronomy is the narration of history.  He enjoyed
engaging his close friend Huib Wouters (see Pictures 22 and 23) in
discussions of Huib's topic, the post-war period in the Netherlands.
Jan and Oskar often wrestled with philosophical and historical issues.
Jan was also attracted to Wittgenstein's philosophy.

Jan's taste in literature was somewhat pessimistic and philosophical.
He particularly enjoyed the novel ``De donkere kamer van Damocles'' by
W.F. Hermans and novels by Celine and Umberto Eco.  He had discovered
the works of the little known Russian absurdist author Daniil Kharms.
In the last years of his life he was reading the series of novels by
J.J. Voskuil that satirized Dutch academic life.  He had the Cambridge
Medieval History on his shelves and many other books on history,
especially the history of science.  He was very appreciative of
nineteenth century publications, such as the beautifully illustrated
works of Camille Flammarion.  He was fascinated by the history of
cartography and read accounts of the Vinland Map intently.  Jan's
interest in language was equally strong.  If I remember correctly, at
one point in the 1980s he was meeting regularly with a group to
decipher the Almagest from the Greek.  He was constantly playing with
words and got a big kick out of the absurdities in the book
``Opperlandse Taal \& Letterkunde'' by Battus from which he could
quote countless nonsense sentences by heart.  He was always searching
for the word in English or Dutch with the longest continuous sequence
of vowels.

One thing that did not interest Jan was computers.  For years he kept
an antiquated Macintosh with a black and white screen at home in
Amsterdam and was irritated and somewhat at a loss whenever the
university system forced him to change his telephone e-mail connection
from home.

\section{GRB 970228} 
The scientific highlight of Jan's life was, no doubt, the discovery of
the optical afterglow of GRB 970228. Govert Schilling captures some of
this fascinating drama and the international political explosion (no
pun implied) that followed in his book {\it Flash!}.  However, he was
not aware of certain tensions that evolved in Jan's group. In the
morning (Holland time) of February 28, Titus Galama (one of Jan's
graduate students) received a message from Jean in 't Zand that a
gamma-ray burst (GRB) had been detected with the Italian satellite
{\it BeppoSAX}. Titus called Jean who gave Titus the position which at
that time was only known to an accuracy of $\sim$10 arc-minutes which
was not good enough to start radio observations at Westerbork.  Titus
called Jan (in Huntsville) to inform him, and Titus kept in touch with
Jean who was working with the {\it BeppoSAX} data to obtain
a refined position with an accuracy of $\sim$3 arc-minutes. When this
refined position became known near 5 PM, Jean told Titus that a
decision had been made higher up in Italy not to release the refined
position to Jan's group. Titus called Jan with the sad news, and Jan
went into high gear. Near 11 PM Amsterdam time, Jan had obtained the 3
arc-minute position from Enrico Costa in Italy.  Jan immediately
called Titus (at home in Amsterdam) who then called Paul Groot who was
still at the Astronomical Institute in Amsterdam (Paul was also one of
Jan's graduate students). It just so happened that Jan's group had
time on the William Herschel Telescope at La Palma that very night
(something that Jan had forgotten, but Paul remembered) to look for a
possible afterglow of GRB 970111.  Paul tried to call Jan to obtain
his approval to take a peek at the position of 970228 in the optical,
but he could not reach Jan. He got on the phone again with Titus, and
they decided to go ahead anyhow without Jan's approval. Paul called
John Telting at La Palma and directed him to observe 970228.

\begin{figure}[!t]
\plotfiddle{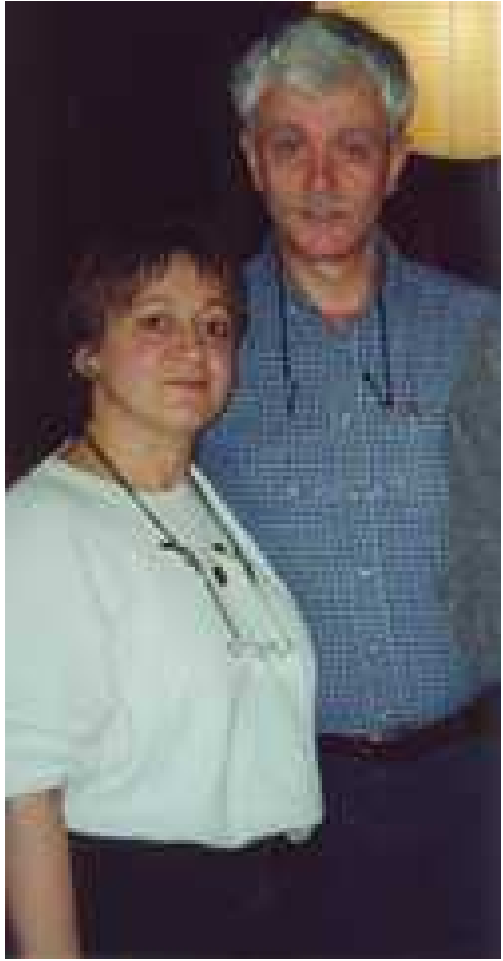}{4.9in}{0}{45}{45}{-105}{-10}
\caption{Jan and Chryssa Kouveliotou at Martha's Vineyard
(May 1998).}
\end{figure}

Given Paul's important contributions (i) to remember that they had
telescope time that very night at La Palma, (ii) to direct John
Telting at his and \mbox{Titus'} initiative to make the optical observations
of 970228 without Jan even knowing this, (iii) to perform the data
analysis, and (iv) to discover the \mbox{afterglow}, Jan agreed that Paul be
first author on the discovery paper of the afterglow. However, when
Jan told me this on the phone (near March 12) I questioned his
wisdom. It's true that without Paul they would have missed the
discovery, but the same could be said of Jean, Titus, and Enrico. My
reasoning was simple: Jan had started a program to search for
afterglows with the Westerbork radio telescopes as early as
1993/94. He thereby had created a framework in which this discovery
could be made, and I therefore argued that he (Jan) should be first
author.

I spent three days in Huntsville (March 16-19) where I gave a
colloquium on kHz QPO; afterwards we had dinner at the home of my
friends Richard and Anna Lieu; Richard is also a professor at
UAH. During my visit, Chryssa and I convinced Jan to undo his promise
to Paul. This was very painful for Jan, but he did. Paul was furious.
However, when I discussed this with Paul recently at MIT, he agreed
with my reasoning though he emphasized that he would certainly not
have agreed with it in 1997. Furthermore, he is still of the opinion
that Jan should not have broken his promise.

\begin{figure}[!b]
\plotone{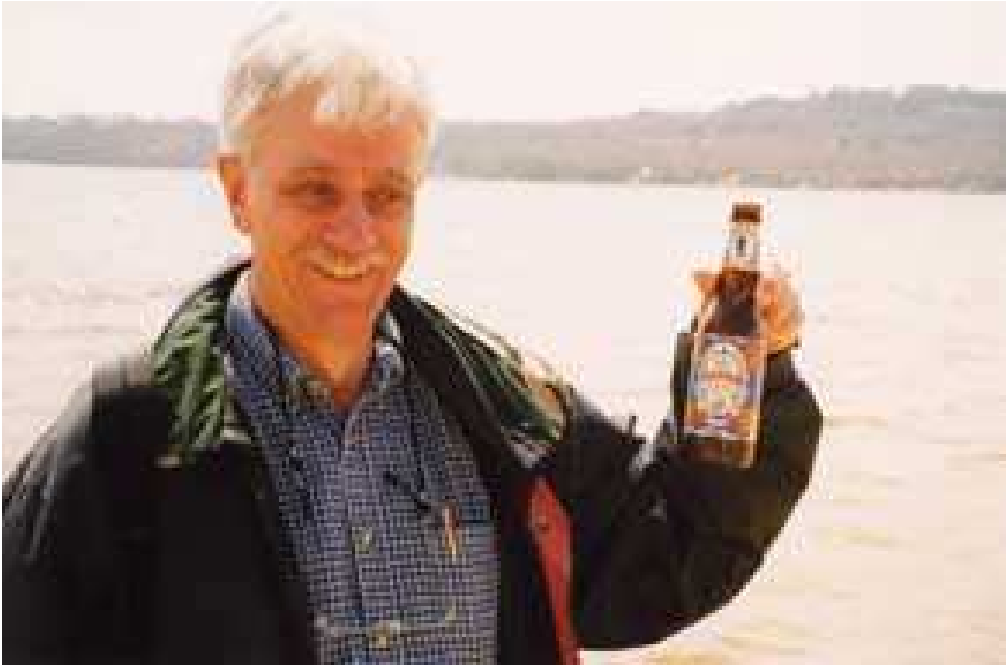}
\caption{Jan's last visit (May 1998). We went to
Martha's Vineyard the weekend before his colloquium on gamma-ray bursts 
at MIT.  He found a beer bottle at Menimsha Pond. He was
very pleased as the ``Harpoon'' label was a first in his large beer
label collection.}
\end{figure}

\begin{figure}[!t]
\plotfiddle{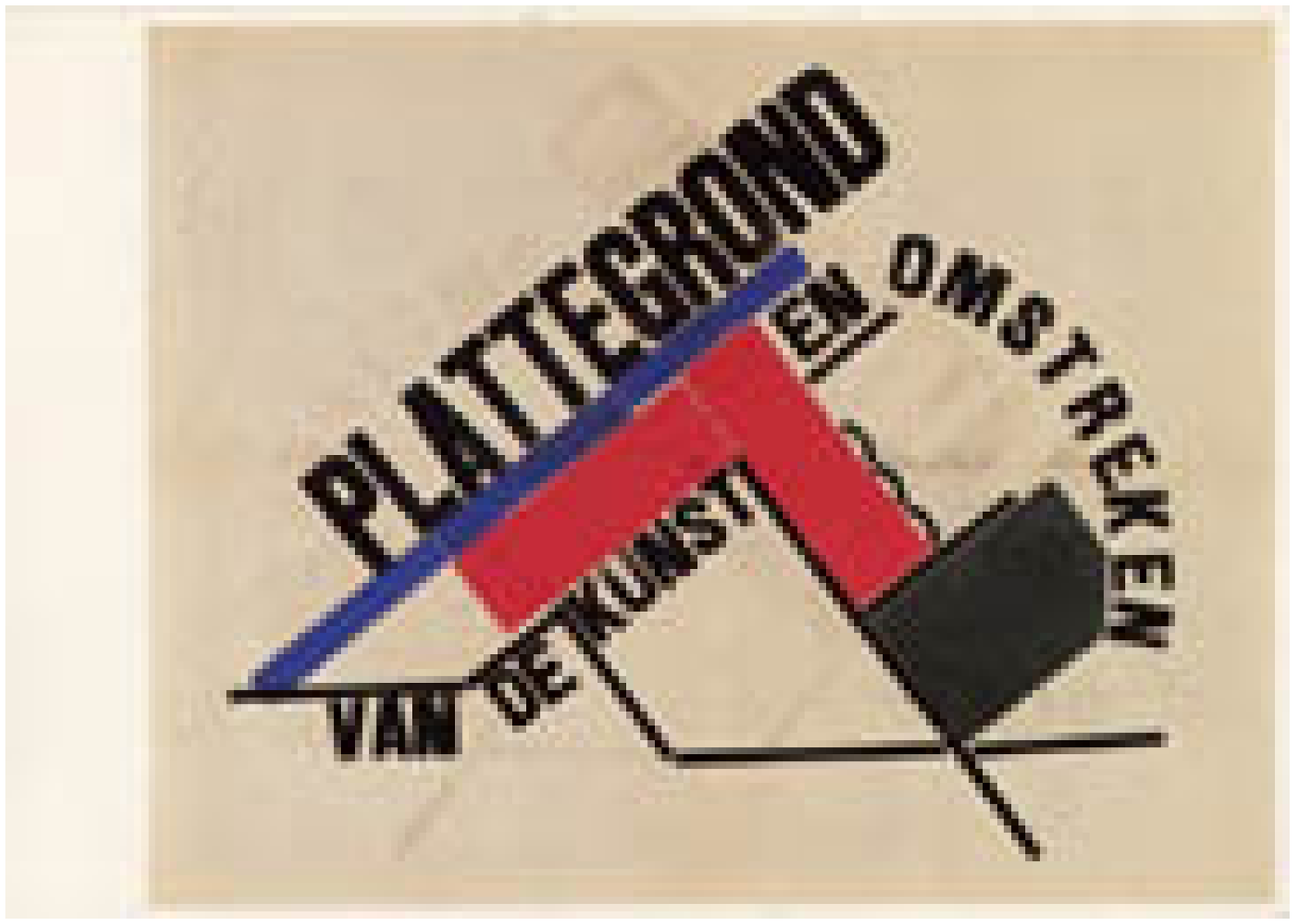}{2.75in}{0}{52}{52}{-160}{0}
\plotfiddle{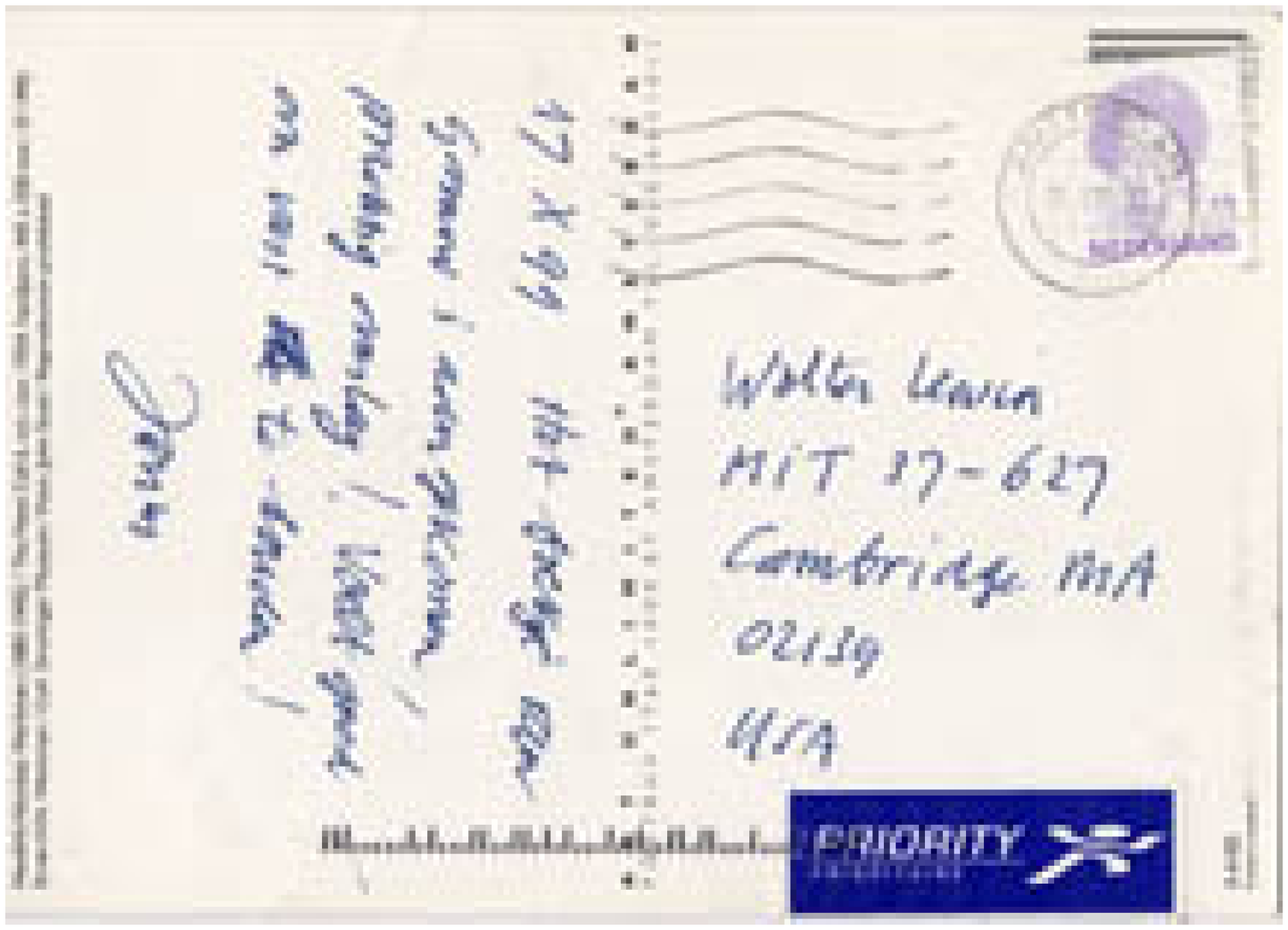}{2.75in}{0}{52}{52}{-160}{-10}
\caption{Jan's last postcard ``The Next Call'' sent to Cambridge on
October 17, 1999, two weeks before he died. The text is encoded, but
I understood what he wanted to say. The stamp of the queen of Holland was designed by
my Dutch friend the artist Peter Struycken.}
\end{figure}

Several people who were also active in the search for afterglows of
GRBs have argued that Jan was lucky. Prior to the detection of the
afterglow of 970228, as early as January 1993, optical searches for
afterglows (with excellent sensitivity) of several GRBs were made
within 24 hours of the occurrence of the GRB (some optical
observations with LOTIS and ROTSE were made within seconds of the
occurrence of the GRB), but no afterglows were observed.  In
hindsight, this is not surprising as we now know that only $\sim$50\%
of all GRBs exhibit an optical afterglow. What is even more remarkable
is the fact that the Galama-Groot-Van Paradijs ``service proposal'' to
search for an afterglow of GRB 970111 was awarded 3 hours observing on
each of two nights on the William Herschel telescope at La Palma, and
those two nights happened to be {\it February 27 and 28, 1997!}  In
addition, it so happened that the position of 970228 \mbox{(Dec
$\sim+12^{o}$, RA $\sim$5h)} was observable from La Palma {\it at
night}, and there were {\it no clouds}. So, yes, everything worked out
``just right''. However, those that call this luck should remember
that Jan's group had built up the necessary expertise over the years,
and they were therefore able to take advantage of this opportunity in
an impressive and decisive way. Given the complexity of the
communications, and the viscous forces at work (rivalry), it is
remarkable that they pulled it off!

Later that year, the distance scale to GRBs was firmly established by
means of redshift measurements of optical afterglows, and Jan and the
{\it BeppoSAX} team were awarded the 1998 Rossi Prize for their
break-through contributions. Jan also received the Physica Award of
the Netherlands Physical Society.

In April 1997, Chryssa gave a seminar at MIT, and Jan and she stayed
with me for five days. That year we also attended a meeting in
Williamsburg and the AAS meeting in Winston Salem, and I stayed with
Jan in Huntsville during the GRB meeting in September.

\begin{figure}[!b]
\plottwo{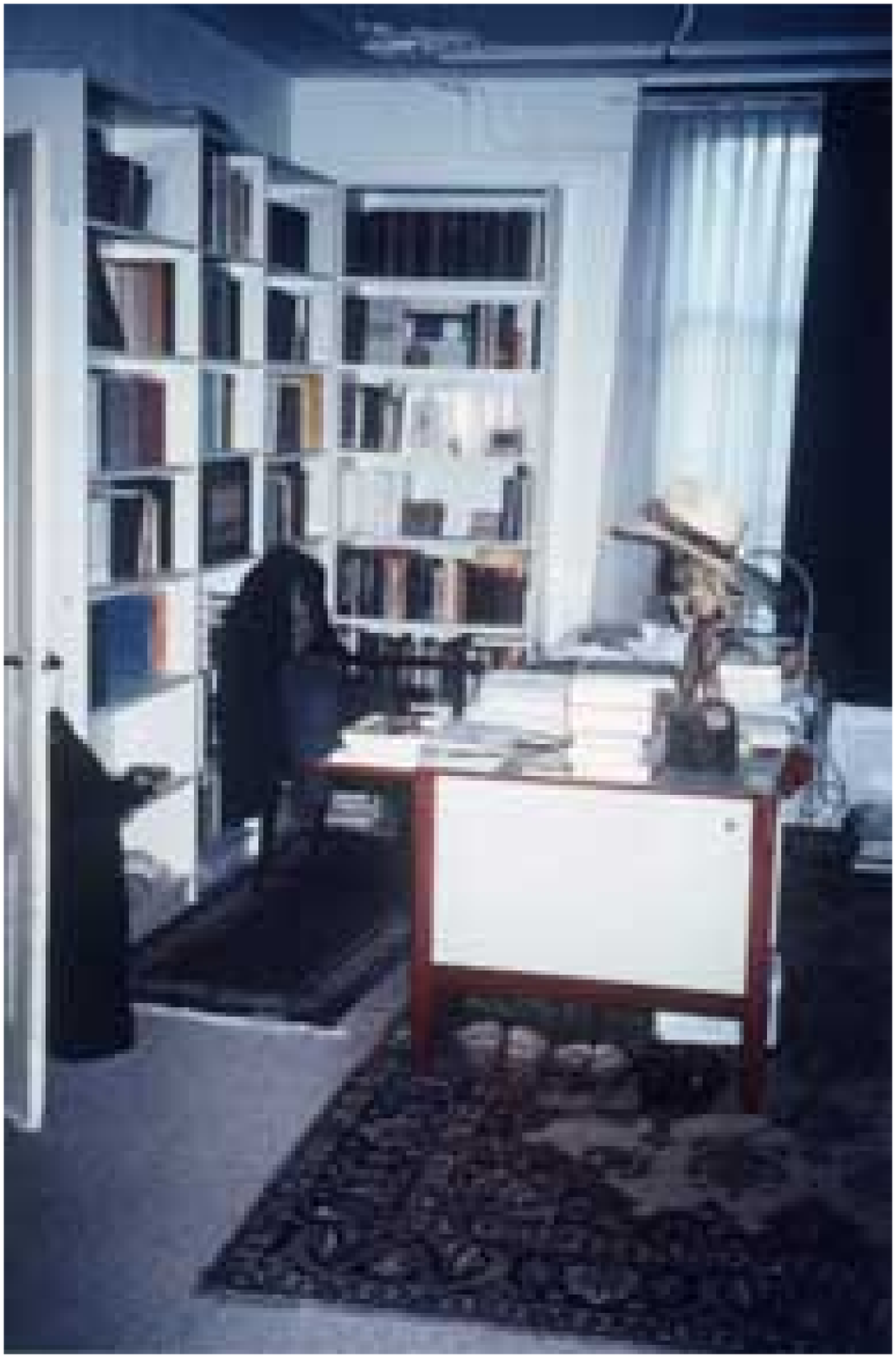}{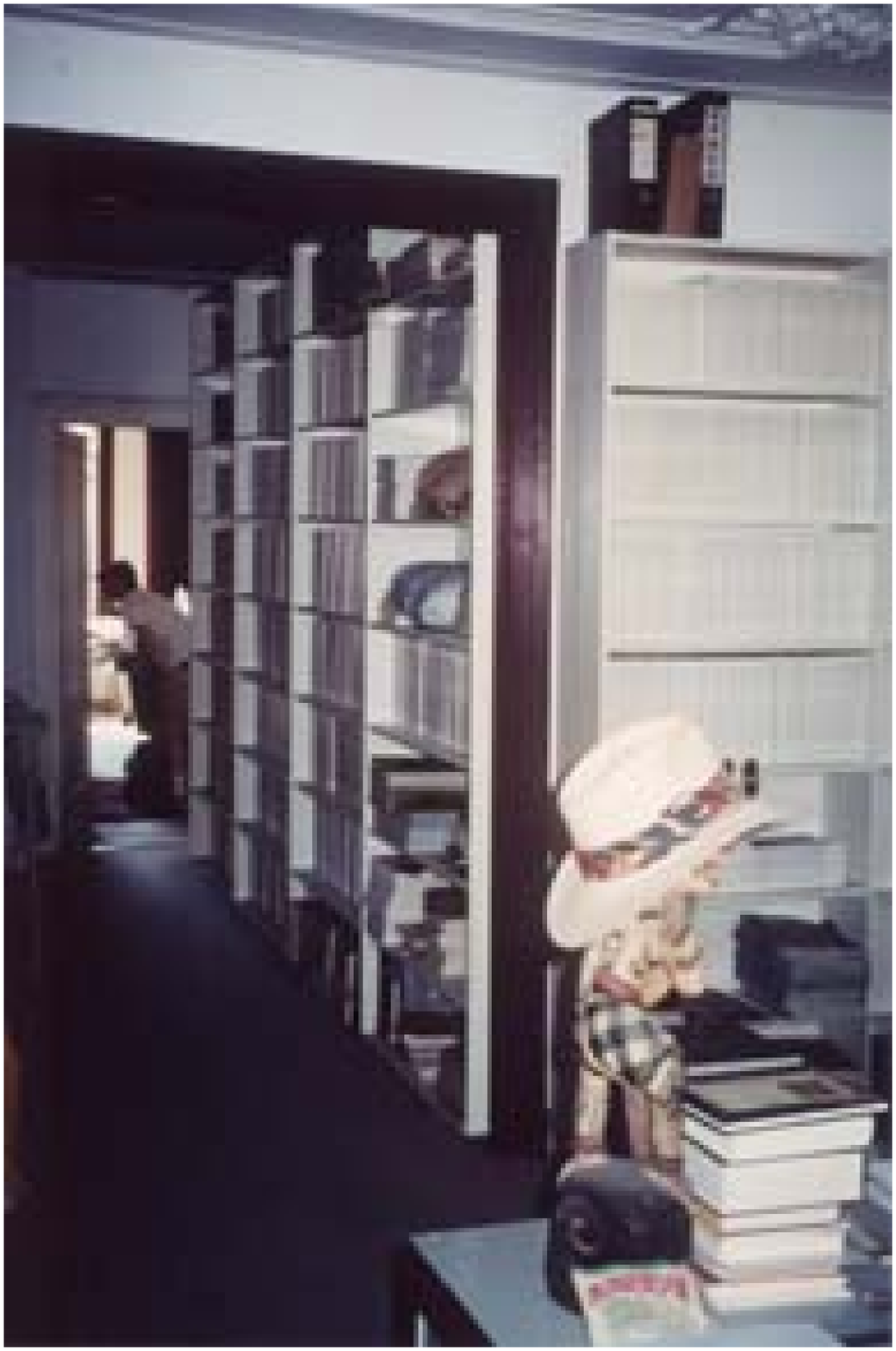}
\caption{{\it Left:} Jan's desk at the Geldersekade in Amsterdam (July
1999). {\it Right:} Some of Jan's astronomical journals.}
\end{figure}

\section{The Last Years} 
The last three years of Jan's life took place in an accelerated
fashion. There was more excitement, happiness, turmoil and sadness in
his scientific and personal life than most people experience in a
life-time.  As far as science is concerned, I already mentioned GRB
970228, but there was more to come.  Chryssa's dedication to Soft
Gamma-ray Repeaters (SGRs) paid off.  Though she deserves most of the
credit, Jan told me that he played an important role.  It is fair to
say that they were the first to demonstrate that the magnetar model
for SGRs, earlier suggested by Duncan and Thompson, is probably the
correct model. This major accomplishment, in my view, warranted the
Rossi Prize, and I am sure that I was not the only person who
nominated Chryssa (and Duncan and Thompson) for this prestigious
award. However, they did not get it (yet).

The year 1998 became extra special for Jan. However, tragedy struck,
and forced by circumstances, his life took a very different course.
Jan had been complaining of a pain in his shoulder starting early
'98. In the fall of that year he mentioned that he had experienced all
of a sudden an excruciating back pain, and he started joking about
``old age''.  He had good weeks and not so good ones; we often talked
about this.

\begin{figure}[!t]
\plotone{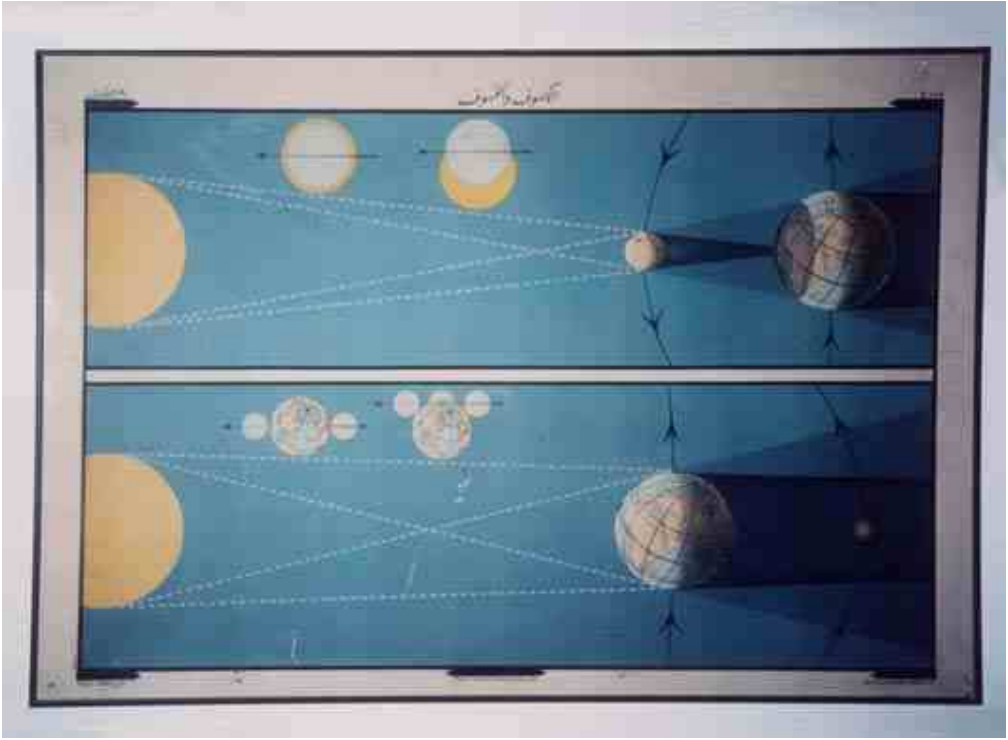}
\caption{A wonderful present from Huib Wouters displayed in
Jan's apartment in Amsterdam (see also picture 23).}
\end{figure}

He stayed with me a few days in January; he gave a colloquium (on GRB)
at Harvard on January 29 (my birthday), and on May 7, he gave a
colloquium at MIT.  We spent the weekend before on Martha's Vineyard
in the house of George Clark on
Edgartown Pond (Pictures 18 and 19). In September that year (during
the Northwest Airline strike) I stayed a few days with Jan in
Amsterdam; this was the last time that we were together at the
Geldersekade.

\mbox{1999} became a disastrous year.  In early February Jan often had a low
grade fever. In February and March he had several tests which showed
that he had {\it spondylolisthesis}, a displacement of the
vertebrae. Meanwhile he was complaining of flu-like symptoms; the low
fever wouldn't go away.  All these complaints: the shoulder, the
fever, the back problems seemed to be unrelated until March 17 when
his doctor in Huntsville came up with {\it multiple myeloma} (a death
sentence) as the diagnosis.  This was confirmed on March 19 but later
turned out to be the wrong diagnosis. Jan had {\it Carcinoma of
unknown primary} which too is terminal.  On the 24th of March we also
learned that there was a large tumor in his liver.  I went to
Huntsville on March 26; Jan was in the hospital.

\begin{figure}[!b]
\plotone{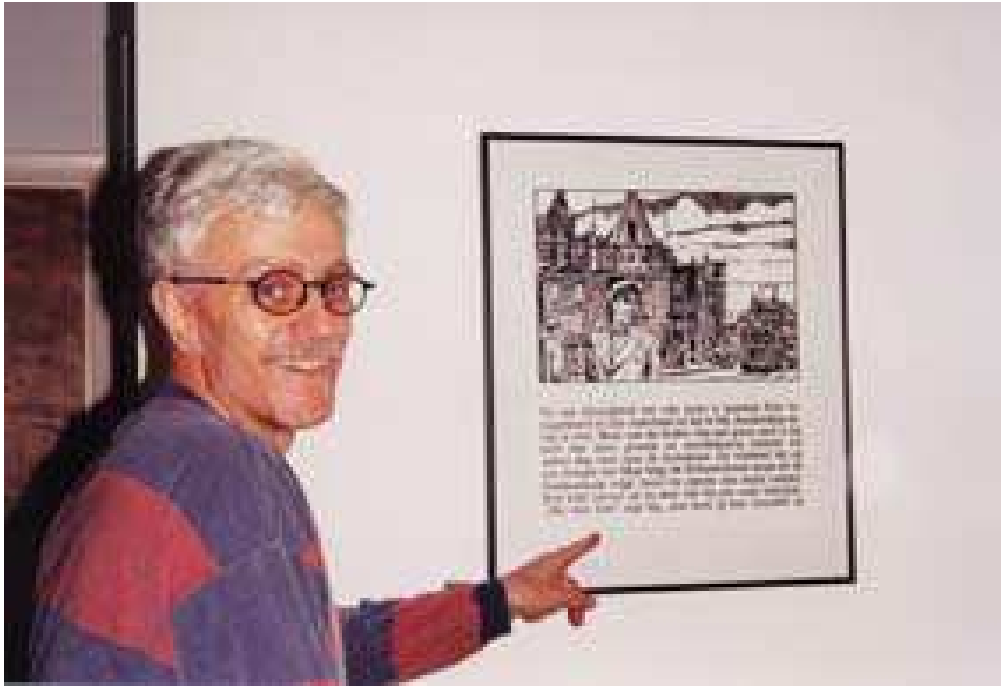}
\caption{Jan in Huntsville pointing at another great
present from Huib Wouters.  It comes from the comic book
about ``Captain Rob'', and makes reference to the Geldersekade and to
``uncle Jan'' {\it \ldots En wanneer hij [kapitein Rob] op een morgen
met Skip langs de Schreierstoren loopt en de Geldersekade volgt, hoort
hij opeens zijn naam roepen. Rob kijkt verrast op en daar ziet hij een
oude bekende. ``H\'{e}, ome Jan!'' zegt hij, ``hoe kom jij hier
verzeild?''.}  This picture was taken in Huntsville by Oskar in April
1999. Photo courtesy Oskar van Paradijs.}
\end{figure}

From that day onward, with very few exceptions, I was every day in
telephone contact with Jan till the very end. At first we were afraid
that in Huntsville Jan would not get the very best cutting edge
treatment possible. Chryssa sent me three protocols for clinical
trials and asked for help in figuring out which was the best way to
go. I forwarded her requests to a friend who was acquainted with
several leading cancer experts in the US and the Netherlands.  In the
months that followed I kept Chryssa informed about their advice every
step along the way. These experts confirmed more than once that at
each stage Jan's doctor(s) in Huntsville chose one of a few ``best''
available options. However, it was a lost battle, and we all knew
that. Jan knew that too, but, understandably, there was always the
hope for a miracle. In fact we all started to believe in a miracle
when Jan's doctor told him on June 18 that the CEA level in his blood
had taken a dramatic dive; it was way down to 30.6! This seemed to
indicate that the latest chemotherapy cure had been successful.
However, that turned out not to be the case. I was in Huntsville June
15--17; this was the last time I saw Jan.

\begin{figure}[!t]
\plottwo{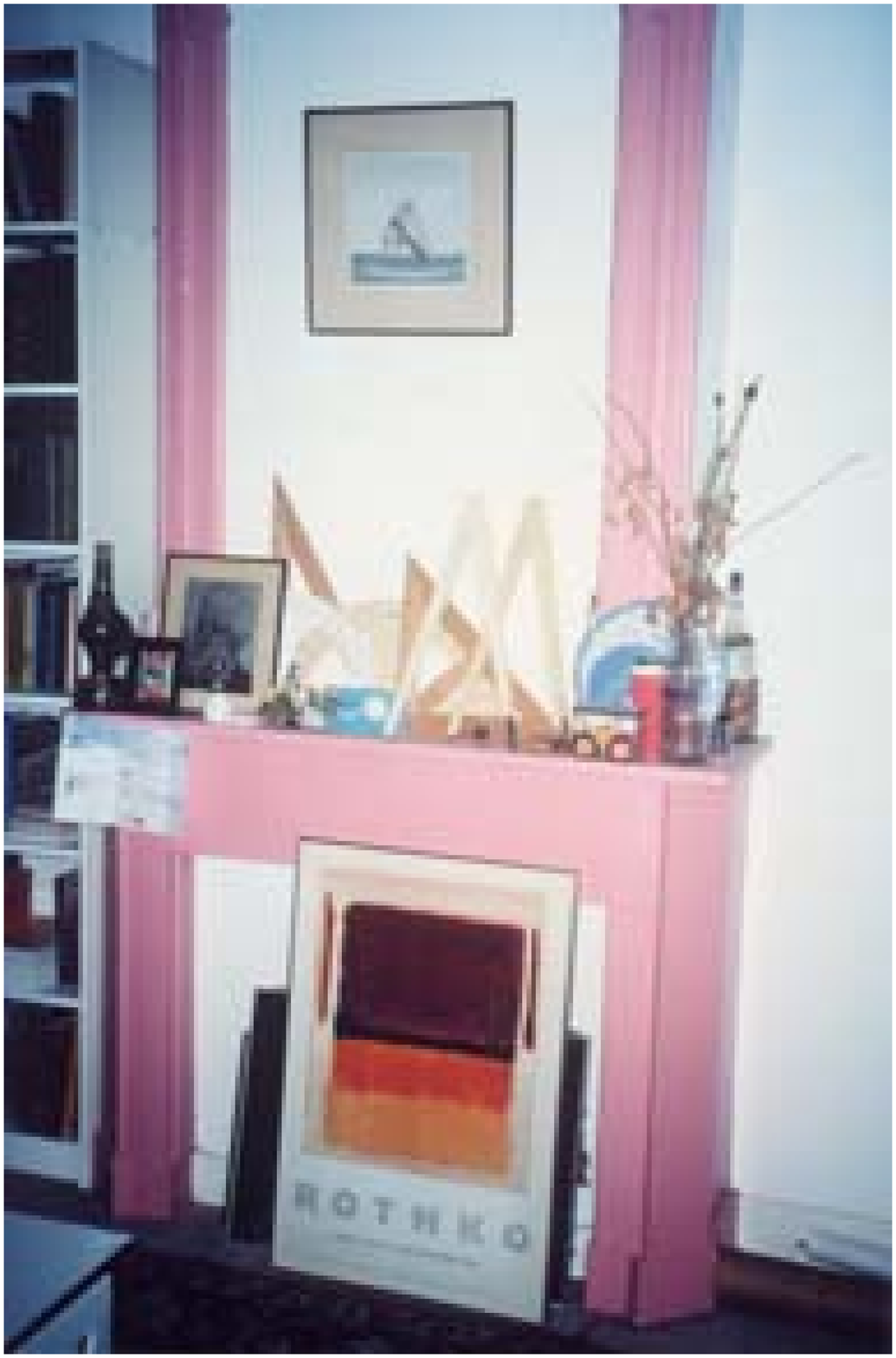}{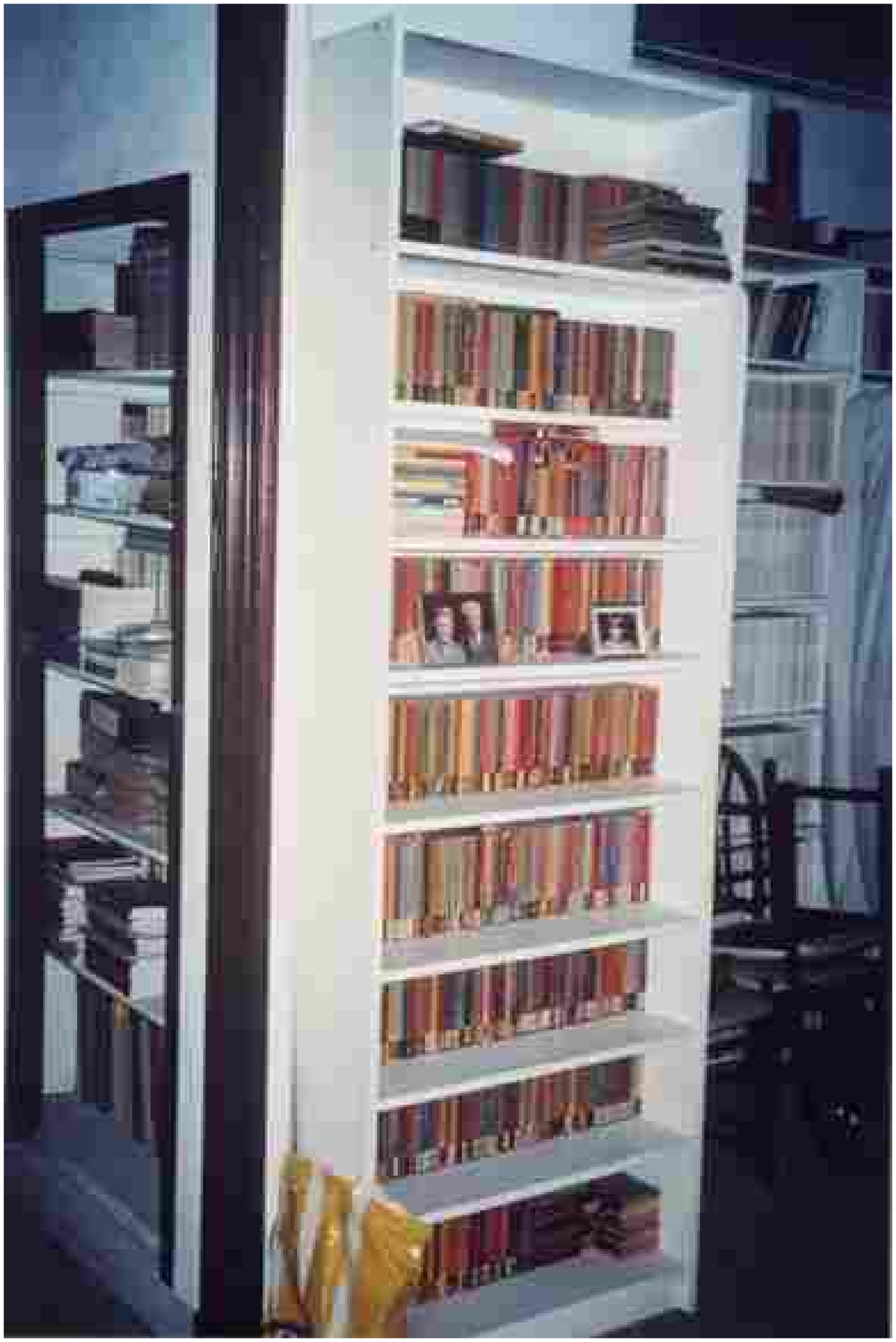}
\caption{{\it Left:} Across from Jan's desk in Amsterdam, a Rothko
which had a very special meaning for him. The framed photo on the left
of the mantle is of Jan's great-grand father Johannes van
Paradijs. Jan's full name was Johannes Antonius. The dry flowers on
the right of the mantle are 15 years old; they were given to Jan by
his mother in 1984 shortly before she died (see text).  {\it Right:}
The book case with Jan's, almost complete, collection of the first 500
Prisma Books. The photo on the left in the bookcase is of Jan's mother
and father; the photo on the right is of his daughter Deirdre.}
\end{figure}

In talking to Jan, I often sensed denial. As an example, Jan suggested
that I spend my next sabbatical in Amsterdam in the fall of 2001, not
in the spring of 2002 as he would then be in Huntsville. However,
there were also times that I talked with Jan openly about the
inevitable. I urged him to write a will (he did not have one), I asked
him where he wanted to die (and be cremated), in Holland or in the US.

After the doctors in Huntsville had given up on Jan (they were not
willing to continue more treatment), Jan went to the Netherlands on
August 26. Ed van den Heuvel had contacted a leading oncologist there
who was willing to continue treating Jan. In Holland Jan became rather
restricted as he could no longer call the US from the hospital.  From
Huntsville he called Cambridge often (sometimes even three times per
day) but that was not possible anymore from the hospital in Amsterdam.

In spite of the serious overall situation, Jan had some nice times in
Holland. He went several times to the Astronomical Institute in
Amsterdam, and he made a trip with Oskar to one of his favorite
antiquarian book stores, De Kloof. He wined and dined at the home of
Ed and AnneMike van den Heuvel, and he visited the Groninger Museum
where he saw the exhibit of my other close Dutch friend Peter
Struycken. On October 17 he sent a last postcard to Cambridge entitled
``The Next Call'': {\it ``Het boekje van Gamow 's aangekomen, prachtig
omslag! Voelt goed om vast te houden! Janus''} (Picture 20). The text
is in Dutch and encoded. Of course, I understood what he really wanted to say.

I last talked to Jan on October 29. On October 30 he was asleep when I
called. On October 31, Chryssa sent me e-mail: ``Jan is slowly
drifting away....''.  I talked to her and to Oskar on November 1; they
told me that Jan would not wake up anymore. He died the next morning,
November~2 shortly after 5 AM (November 1, 11 PM in Cambridge).

Over the years I have stayed countless weeks at the Geldersekade;
often together with Jan and later also alone or with a friend when Jan
was in Huntsville or on vacation in Greece. When I stayed there with
Susan Kaufman in July 1999, I
realized that this was going to be my last visit, and I made a series
of slides to remember Jan's home with so many memories (Pictures
21, 22, 24, 25).

\section{Our Special Bond}
Science had brought us together and science continued to play a
central role throughout our lives. We often bounced scientific ideas
around (Jan had so many) by e-mail or by phone or in person --- we
spent thousands of hours writing proposals and articles together and
we worked relentlessly for years on ``our'' book {\it X-ray Binaries}
which Jan and I called {\it the bible} (my students still do).  Who
was first or last on a publication was never an issue between us.
However, Jan once (in 1983) asked me to be last on a paper which had
52 authors. Jan reasoned that being last was more prestigious than
being the 3rd author (Andy Lawrence and Lynn Cominsky were first and
second author). Helmut Abt told me later that because of this paper
the Astrophysical Journal changed its policy of indexing authors.
Before the policy change, the space needed for indexing scaled
approximately as the square of the number of authors. In 1992, Jan and
I were approaching our 100th paper together, and Jan suggested that we
carefully plan (and delay) our Space Science Review on X-ray bursts
(with Ron Taam) so that it would become exactly our hundredth paper,
and it did!

\begin{figure}[!t]
\plotfiddle{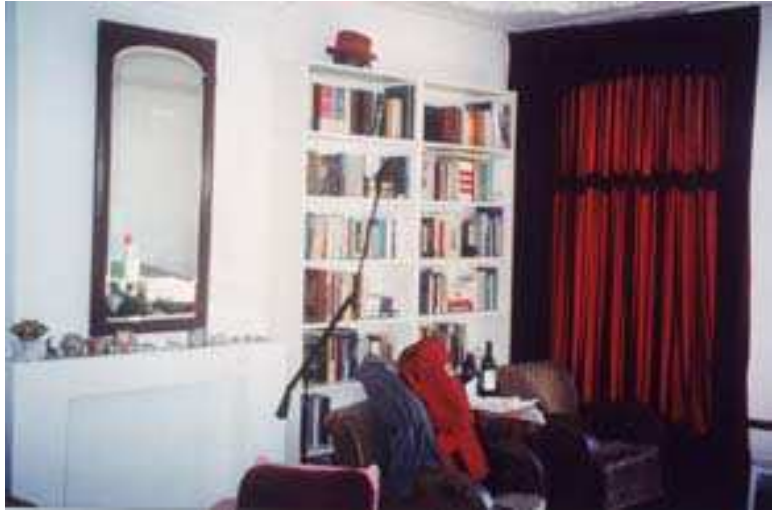}{2.75in}{0}{52}{52}{-160}{-10}
\plotfiddle{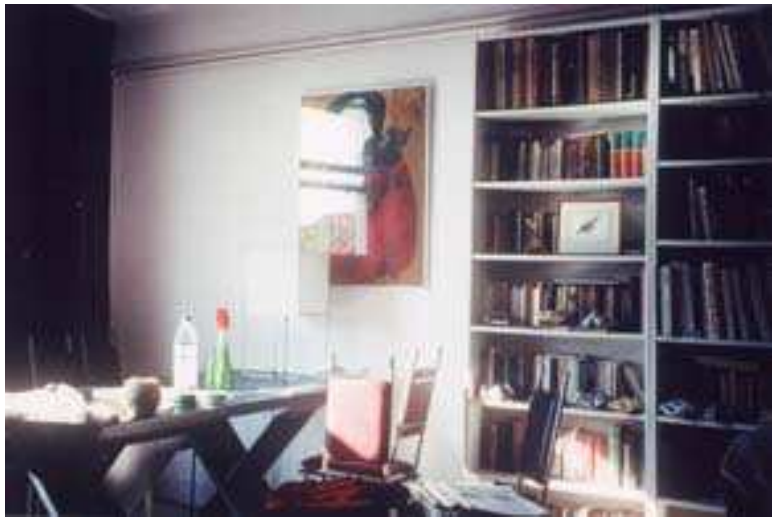}{2.75in}{0}{52}{52}{-160}{-10}
%\plottwo{jvp28.ps}{jvp27.ps}
\caption{Two views of the
living room of Jan's apartment at the Geldersekade. On the wall
(bottom picture) a work of art by Jan's brother Ren\'{e}; during
the last months of his life Jan verbally (unfortunately not in writing) willed this to me. When
looking through the windows (the curtains are closed here), one can
see the activity across the street in the red light district (see
text).}
\end{figure}
Jan had a photographic memory for references. Prior to the age of
e-mail, when I needed a reference I would call Jan in Amsterdam. He
was absolutely amazing. All I had to do was give him a few clues and
more often than not, he would mention with great confidence the
journal, the volume and even the page number. Such calls would take
less than one minute; much less time for me than to go to the library.
Jan did not mind at all that I bothered him frequently with such
mundane questions. On the contrary, he took great pride in
demonstrating his encyclopedic memory for references.

It wasn't only the science that was important in our lives. Our
personal bonds and deep friendship grew over the years: we shared the
good and the bad, victory and defeat, the highs and the lows; we
laughed and cried together, we lied for each other, and during times
of immense pain we were always there for each other.  He called me
{\it Wout} and I called him {\it Janus} (see the text of Pictures 20
and 26). Whenever I needed support in my life during down periods (and
there were several), Jan would call often to check on how I was doing,
and he would make every effort to get me back on my feet. I too
supported him to the best of my ability during his difficult
times. During the later part of his life, Jan developed a rather
philosophical attitude towards the many episodes that caused great
stress. He would say: {\it het regent} (it's raining).

Over the years we developed a way of communicating that was almost
like a secret language which no one else could possibly understand.
An example is the text of Jan's last postcard (Picture 20). We also
mixed English with Dutch to express emotions and feelings. We would
use quotes that had special meaning only to us.  Jan and I were very
much alike; we were like brothers and resonated in many ways. We could
explode in laughter in the presence of others who did not have a clue
why we laughed so uncontrollably.  This was not always polite and it
once, in 1997, caused a major problem.

Jan sent me a very special card on my 63rd birthday in January 1999
(Picture 26). It shows two boxes similar in shape to the lamps by the
Dutch designer Benno Premsela (you can see one in Picture 15 behind
Jan).  The text reads: {\it Wout, ik wens je toe dat de transformatie
je heel goed zal bevallen! Janus.} On the side he wrote: {\it krijgt
het komend jaar zijn kleuren wel . . . .} Freely translated: {\it
  Wout, I
hope that the transformation will be a pleasant one!  Janus}. On the side:
{\it will get its colors in the year to come}.  At first I wondered
what this riddle was all about. But I soon got it.  This was so
typical Jan; the choice of the Premsela lamps also had a very special
meaning! One of Jan's graduate students Peter Woods, in Huntsville,
wrote me about the card the following: ``We were writing papers and
proposals and the proposal deadline was nearing. I had a question for
Jan so I went to his office.  He was sitting at his desk which was
cleared except for the card and a pile of colored markers.  My first
thought was Dear God, he's lost it now, he's reverted back to
coloring! Of course, I didn't say this out loud.  I asked him what he
was doing and he said that he was making you a birthday card.  He
showed it to me and translated the text {\it Enjoy the
Transformation}.  I couldn't figure out what he meant by the colored
boxes, so he explained that there were 63 cubes on the left and 64 on
the right to indicate your change in years. I enjoyed the card; it was
very clever. What impressed me about Jan is that in spite of his
hectic schedule and the deadlines we were facing, he cleared his desk
and spent time to write his best friend a birthday card.''

\begin{figure}[!b]
\plotone{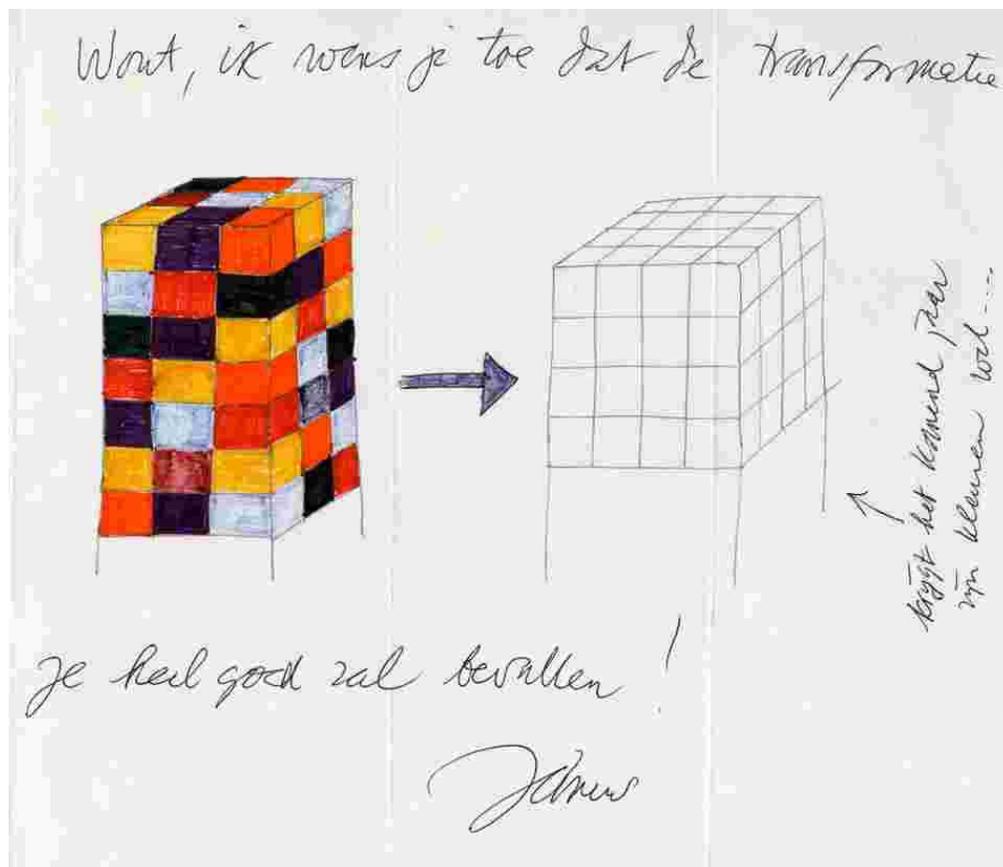}
\caption{The card that Jan made for my 63rd birthday 
in January 1999 (see text).}
\end{figure}

Jan's apartment at the Geldersekade 91$^{3}$ was located in the red
light district in the very center of Amsterdam, close to the Central
Railroad Station, the Nieuwmarkt, and the Schreierstoren (see Picture
23). There is no safer place to live in Amsterdam, since police
officers can be found at every corner. From Jan's apartment one could
easily see how well the ``business'' across the street was doing, and
we once made some measurements of the amount of time that the various
clients stayed, and the average time between customers.  This was
quite easy as all we had to do is to time the opening and closing of
the window curtains across the street (Picture 25).  We even
made QPO-like plots of ``frequency'' vs.\ time.

Neither Jan nor I were heavy drinkers, but what was always very
special for us was a 17 year old Dutch ``jenever'' that was sold (\$80
per bottle) at only one liquor store in Amsterdam.  Drinking this
stuff was almost a ritual. We would have one (or two) before bed. Jan
used to say: ``how about a pikketanisje,'' and I always knew what was
coming.

Jan was the executor of my will; we agreed that it was only reasonable
and fair for me to die first as I was ten years older. I willed him a
work of contemporary art from my collection. At my request he
hand-picked it himself. His choice (a very low-key work by Peter
Struycken) was quite remarkable, and I felt proud of him for picking
that one.  But nature was cruel, Jan died first, and our journey came
to an abrupt end.  Few people know (perhaps only two) what we meant
for each other. When Jan died, part of me died.

\acknowledgments
I am very grateful to Oskar van Paradijs,
Derek Fox, Jefferson Kommers, Peter Struycken, and two persons who
prefer to remain anonymous, for their valuable comments on this
manuscript and for their very thoughtful suggestions. I thank George
Clark, Lynn Cominsky, Titus Galama, Paul Groot, Jeff Hoffman, Kevin
Hurley, Shri Kulkarni, Erik Kuulkers, Richard Lieu, Eugene Magnier,
Jeff McClintock, Maria Moesman, Tony Peacock, Holger Pedersen, Luigi
Stella, Lidewijde Stolte, Joachim Tr\"{u}mper, Ed van den Heuvel,
Michiel van der Klis, Erica Veenhof, Frank Verbunt, Nick White, Peter
Woods and the Koninklijke Bibliotheek in The Hague for their help and
inputs, and Dave Pooley for scanning the pictures and for his tireless
efforts in formatting this manuscript.  I want to thank Maria Moesman
for her hospitality during my visits to Haarlem, and Chryssa
Kouveliotou for her hospitality during my visits to Huntsville.

\end{document}